\documentclass{aa}  

\usepackage[usenames, dvipsnames]{xcolor}
\usepackage{graphicx}
\usepackage{txfonts}
\usepackage{amsmath}
\usepackage{textgreek}

\usepackage{natbib,twoopt}
\usepackage[colorlinks=true,allcolors=blue]{hyperref} 
\bibpunct{(}{)}{;}{a}{}{,} 
\makeatletter
\newcommandtwoopt{\citeads}[3][][]{\href{http://adsabs.harvard.edu/abs/#3}%
{\def\hyper@linkstart##1##2{}%
\let\hyper@linkend\@empty\citealp[#1][#2]{#3}}}
\newcommandtwoopt{\citepads}[3][][]{\href{http://adsabs.harvard.edu/abs/#3}%
{\def\hyper@linkstart##1##2{}%
\let\hyper@linkend\@empty\citep[#1][#2]{#3}}}
\newcommandtwoopt{\citetads}[3][][]{\href{http://adsabs.harvard.edu/abs/#3}%
{\def\hyper@linkstart##1##2{}%
\let\hyper@linkend\@empty\citet[#1][#2]{#3}}}
\newcommandtwoopt{\citeyearads}[3][][]%
{\href{http://adsabs.harvard.edu/abs/#3}
{\def\hyper@linkstart##1##2{}%
\let\hyper@linkend\@empty\citeyear[#1][#2]{#3}}}
\makeatother
\usepackage{siunitx}
\DeclareSIUnit \parsec {pc}
\DeclareSIUnit\angstrom{\text {√Ö}}
\DeclareSIUnit\year{\text {yr}}
\DeclareSIUnit\erg{\text {erg}}
\DeclareSIUnit\jansky{\text {Jy}}
\DeclareSIUnit \solarmass {\ensuremath{M_\odot}}
\DeclareSIUnit \h {\ensuremath{\mathit{h}}}

\usepackage{ulem}

\newcommand{\Obelisk}{\textsc{Obelisk}}
\newcommand{\fedd}{\ensuremath{f_{\rm Edd}}}

\newcommand{\msun}{\,{\rm M_\odot}}
\newcommand{\mstar}{\ensuremath{M_{\rm star}}}
\newcommand{\mbh}{\ensuremath{M_{\rm BH}}}

\begin{document}

   \title{Exploring Active Galactic Nuclei and Little Red Dots with the Obelisk simulation}

   \author{M. Volonteri \inst{1}        
          \and
          M. Trebitsch \inst{2}
          \and 
          J. E. Greene \inst{3}
          \and 
          Y. Dubois \inst{1}
          \and 
          C.-A. Dong-Paez \inst{1}
          \and 
          M. Habouzit \inst{4,5}
          \and 
           A. Lupi\inst{6,7}
           \and 
           Y. Ma\inst{3}
           \and 
          R. S. Beckmann\inst{8}
          \and 
          P. Dayal\inst{2}  
          \and 
          R. Schneider\inst{9,10,11,12}  
          }

   \institute{Institut d'Astrophysique de Paris, Sorbonne Universit\'e, CNRS, UMR 7095, 98 bis bd Arago, 75014 Paris, France\\
              \email{martav@iap.fr}
         \and
             Kapteyn Astronomical Institute, University of Groningen, PO Box 800, NL-9700 AV Groningen, the Netherlands,
             \and 
             Department of Astrophysical Sciences, Princeton University, Princeton, NJ 08544, USA, 
     \and
     Max-Planck-Institut f\"ur Astronomie, K\"onigstuhl 17, D-69117 Heidelberg, Germany,
     \and
    Zentrum für Astronomie der Universit\"at Heidelberg,
ITA, Albert-Ueberle-Str. 2, D-69120 Heidelberg, Germany,
             \and
            Dipartimento di Scienza e Alta Tecnologia, Universit\`a degli Studi dell'Insubria, via Valleggio 11, I-22100, Como, Italy,
        \and
            INFN, Sezione di Milano-Bicocca, Piazza della Scienza 3, I-20126 Milano, Italy,
\and
Institute for Astronomy, University of Edinburgh, Royal Observatory, Edinburgh EH9 3HJ, UK
\and
Dipartimento di Fisica, ``Sapienza'' Universit$\grave{a}$ di Roma, Piazzale Aldo Moro 2, 00185 Roma, Italy 
\and
INAF/Osservatorio Astronomico di Roma, Via di Frascati 33, 00040 Monte Porzio Catone, Italy 
\and
INFN, Sezione Roma1, Dipartimento di Fisica, ``Sapienza'' Universit$\grave{a}$ di Roma, Piazzale Aldo Moro 2, 00185, Roma, Italy
\and 
Sapienza School for Advanced Studies, Viale Regina Elena 291, 00161 Roma, Italy }


 
  \abstract{The James Webb Space telescope has discovered an abundant population of broad line emitters, typical signposts for Active Galactic Nuclei (AGN). Many of these sources have red colors and a compact appearance that has led to naming them `Little Red Dots'. In this paper we develop a detailed framework to estimate the photometry of AGN embedded in galaxies extracted from the \Obelisk{} cosmological simulation to understand the properties of color-selected Little Red Dots (cLRDs) in the context of the full AGN and massive black hole population. We find that using realistic spectral energy distributions (SEDs) and attenuation for AGN we can explain the shape of the cLRD SED as long as galaxies host a sufficiently luminous AGN that is not too much or too little attenuated. When attenuation is too low or too high, AGN do not enter the cLRD selection, because the AGN dominates over the host galaxy too much in blue filters, or it does not contribute to photometry anywhere, respectively. cLRDs are also characterized by high Eddington ratios, possibility super-Eddington, and/or high ratios between black hole and stellar mass. 
  }


   \maketitle
%

\section{Introduction}
\label{sec:intro}

Over the past year the James Webb Space Telescope (JWST) has opened a new window on the high-redshift Universe, discovering galaxies and Active Galactic Nuclei (AGN) at unprecedented redshifts \citep[see][for a review]{2024arXiv240521054A}. The vast majority of the AGN has been selected or confirmed through broad emission lines, which generally are a signpost for the deep potential well of MBHs \citep[e.g.,][]{2024ApJ...971...68W}. 

A long-standing technique to pre-select high-z AGN candidates is color selection \citep[e.g.][]{2001AJ....121...31F}. The different shape of galaxy and AGN spectral energy distributions (SEDs) modulates the colors of sources, isolating in different regions those that are AGN or galaxy dominated, at least when the ratio between the luminosity of the two components is sufficiently different \citep{2017ApJ...849..155V,2018MNRAS.476..407V}.   
Additionally, at optical wavelengths AGN can also be selected by variability \citep[e.g.,][]{2024ApJ...971L..16H} and by having a compact appearance \citep[e.g.][]{1998AJ....116.2624J}, as AGN appear as point sources, and when sufficiently more luminous than the host galaxy outshine it.  

 Search for broad-line emitters with JWST were performed on targets selected in the JADES survey \citep{2024A&A...691A.145M}, using data from spectroscopy programs \citep{2023ApJ...959...39H,2023ApJ...954L...4K} and in the ASPIRE spectroscopic survey \citep{2024ApJ...974..147L}.
 \citet{2024ApJ...963..129M} performed a search for broad-line emitters in the FRESCO \citep{2023MNRAS.525.2864O} and EIGER \citep{2023ApJ...950...66K} finding an abundant population, characterized by red colors and compact sizes, winning them the name  `Little Red Dots'. \citet{2024ApJ...964...39G} searched in the UNCOVER survey \citep{2024ApJ...974...92B} for compact red sources, and follow-up spectroscopy confirmed that they were broad line emitters.  Many if not most of the Little Red Dots show a blue excess in the short wavelength NIRCAM bands \citep{2024ApJ...964...39G,2024arXiv240403576K}, which has been ascribed to either the host galaxy emission or scattered light from the AGN.  Following these discoveries, other photometric searches were performed, selecting for red sources \citep[e.g.][]{2024ApJ...968...38K,2024arXiv240403576K} and identifying a very large population of sources. If all these red sources are AGN, they would severely challenge our understanding of galaxy and massive black hole (MBH) evolution \citep{2024arXiv240610341A}.

In this paper we want to investigate the properties of MBHs in galaxies to understand the relationship between Little Red Dots and the general population of AGN, as well as what is needed for an AGN to be a Little Red Dot. Specifically, we model and analyze simulated photometry of galaxies and AGN based on the \Obelisk{} simulation, select red sources, to which we will refer as `color Little Red Dots' (cLRDs), and explore their properties.

\section{Methodology}
\label{sec:methods}

\subsection{The \Obelisk{} simulation}
\textsc{Obelisk} \citep{Trebitsch2021} is a high-resolution resimulation down to redshift $z \simeq 3.5$ of the most massive halo at $z=1.97$ in the \textsc{Horizon-AGN} \citep{Dubois2014c} simulation. The simulation is based on a \textLambda CDM cosmology with  WMAP-7 parameters \citep{Komatsu2011}: $H_0 = \SI{70.4}{\kilo\meter\per\second\per\mega\parsec}$,  $\Omega_\Lambda = 0.728$, $\Omega_\mathrm{m} = 0.272$, $\Omega_\mathrm{baryon} = 0.0455$,  $\sigma_8=0.81$, and spectral index $n_\mathrm{s}=0.967$. The zoomed-in region has a volume of $\sim 10^4 \, h^{-3}\,\mathrm{cMpc}^3$. We summarize here the main assumptions, and refer the readers to \citet{Trebitsch2021} and \citet{2023A&A...673A.120D} for further details. 

 \textsc{Obelisk} was run with \textsc{Ramses-RT} \citep{Rosdahl2013,Rosdahl2015}, a radiative transfer hydrodynamical code which builds on the adaptive mesh refinement (AMR) \textsc{Ramses} code \citep{Teyssier2002}. Cells are refined up to a smallest size of $35\,\si{\parsec}$ if its mass exceeds $8$ times the mass resolution. The simulation assumes an ideal monoatomic gas with adiabatic index $\gamma = 5/3$ and includes gas cooling and heating down to very low temperatures ($50\,\rm K$) with non-equilibrium thermo-chemistry for hydrogen and helium, and contribution to cooling from metals (at equilibrium with a standard ultraviolet background) released by SNe. 

In the high-resolution region dark matter particles have mass $1.2\times 10^6\, \msun$, star particles have mass $\sim 10^4\,\msun$, and assume a \citet{Kroupa2001} initial mass function between $0.1$ and $100 \,\msun$. Stars form in gas cells with density higher than $5\,\mathrm{H}\,\si{\per\cm\cubed}$ and Mach number $\mathcal{M}\geq 2$, with a star formation efficiency that depends on the local gas density, sound speed, and turbulent velocity. Mechanical SN feedback is released  $5\,\si{\mega\year}$ after the birth of a stellar particle, with a mass fraction of $0.2$, and releasing $10^{51}\,\si{\erg}$ per SN.  \textsc{Obelisk} also includes modelling of dust as a passive variable. 

MBHs form when in cells where both gas and stars exceed a density threshold of $100 \, \si{H} \, \si{\per\centi\meter\cubed}$ and the gas is Jeans unstable, with an exclusion radius of 50 comoving kpc to avoid formation of multiple MBHs. The initial MBH mass is  $3\times 10^4\,\msun$, and MBHs grow by accretion, capped at the Eddington rate,   
using a Bondi-Hoyle-Lyttleton (BHL) formalism, where the accretion rate is calculated via the local average gas density, gas sound speed, and MBH relative velocity with respect to the gas. 

\Obelisk{} includes a dual-mode AGN feedback. Below $f_\mathrm{Edd}<0.01$ (where $\fedd$ is the ratio of the MBH mass accretion rate over its Eddington mass accretion rate), AGN releases a fraction $\varepsilon_\mathrm{MCAF}$ of the accretion energy as kinetic energy in jets. Jets assume Magnetically Chocked Accretion Flows, and $\varepsilon_\mathrm{MCAF}$ is a polynomial fit to the simulations of \citet{McKinney2012} as a function of MBH spin. At higher $f_\mathrm{Edd}\geq 0.01$, the $15\%$ of the accretion luminosity is released isotropically as thermal energy. MBH spin magnitudes and directions are self-consistently evolved on the fly via gas accretion and MBH-MBH mergers. Two MBHs are merged when their separation becomes smaller than $4\Delta x$. The simulation models dynamical friction explicitly, including both gas and collisionless particles (stars and DM) \citep[following the implementation presented in][]{Dubois2013,2019MNRAS.486..101P}. The dark matter distribution is smoothed by a cloud-in-cell interpolation to avoid artificial scattering.

   \begin{figure}
   \centering
   \includegraphics[width=\columnwidth]{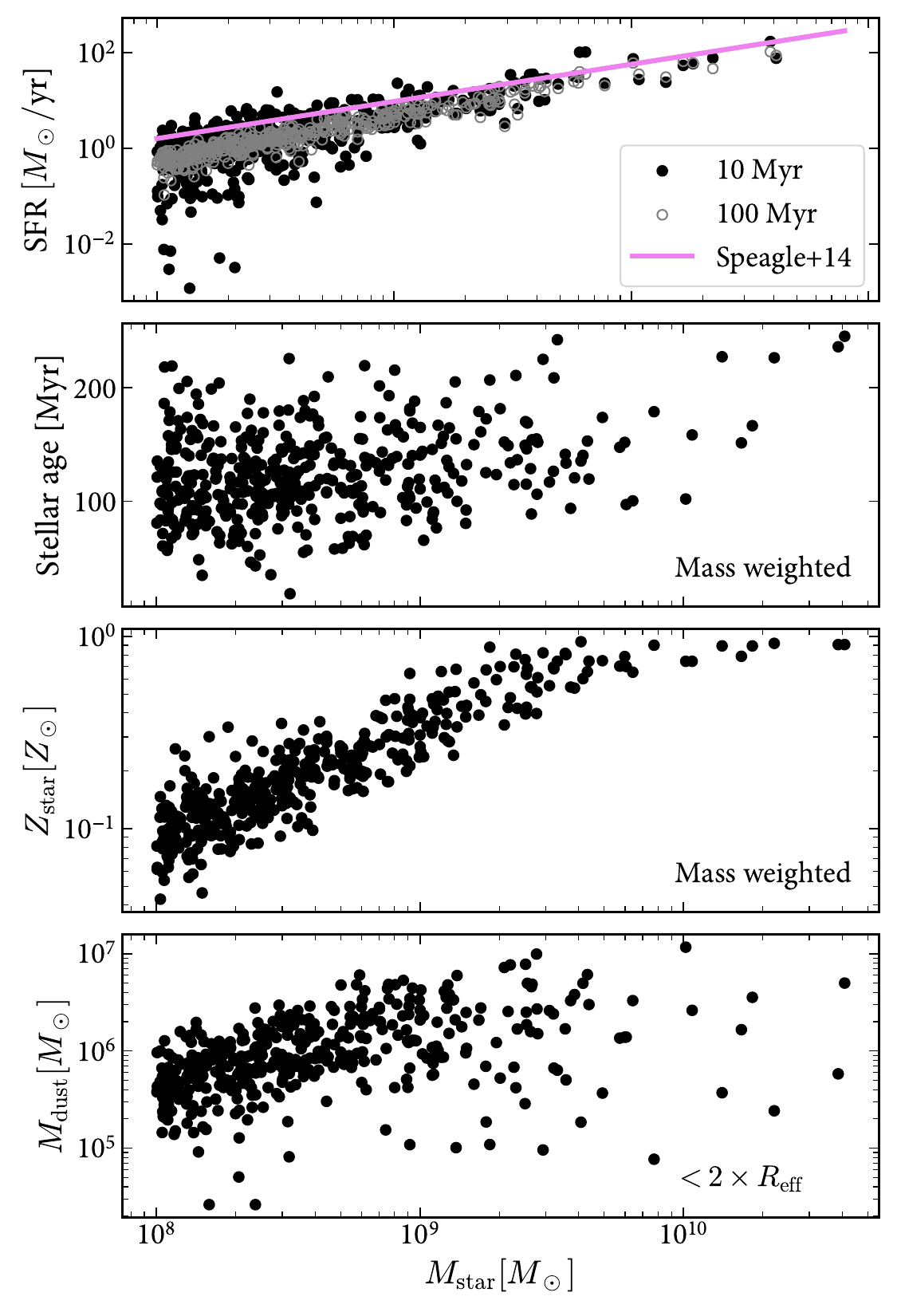}
   \caption{Properties of  simulated galaxies with $\mstar>10^8 \msun$ at $z=6$, including galaxies that do not host MBHs. From top to bottom: star formation rate averaged over 10~Myrs in black, and over 100~Myrs in gray; mass-weighted stellar age, mass-weighted stellar metallicity, where solar metallicity is assumed to be 0.02, dust mass within twice the effective radius. $R_{\rm eff}$ is defined as the half-mass projected radius in the plane orthogonal to the stellar angular momentum, computed as the (2D) radius such that 50\% of the mass is within that radius. The violet line in the top panel shows the fit from \citet{2014ApJS..214...15S}.}
              \label{fig:galprops}
    \end{figure}

 \Obelisk{} includes a model for dust evolution. The model assumes that dust grains are released in the ISM via SN explosions, grow in mass via accretion of gas-phase metals, and are destroyed by SN explosions and via thermal sputtering. Dust grains are further assumed to be perfectly coupled to the gas, so that they can be described by a scalar passively advected like the metallicity. We assume that all grains have an average size of $a_{\rm g} = 0.1\,\mu\mbox{m}$ and density $\mu_{\rm g} = 2.4\,\mbox{g}\,\mbox{cm}^{-3}$. Further details can be found in \citet{Trebitsch2021} and \citet{2024A&A...687A.240D}.

The galaxy and AGN populations in \Obelisk{} are presented in \citet{Trebitsch2021}. Galaxies are consistent with the expected relation between stellar and dark matter halo mass, and the star formation rate density is in good agreement with that found in overdense regions. Galaxy masses are also not challenged by JWST results \citep{2023ApJ...943L..28K}. The simulation broadly produces gas, metals, and dust content in high-z galaxies in agreement with observed scaling relations with stellar mass (Trebitsch et al. in prep). The properties of the galaxies analyzed in this work are shown in Fig.~\ref{fig:galprops}. AGN in \Obelisk{} are abundant but MBHs mainly grow only in galaxies with $\mstar>10^9 \, \msun$ because of the effect of SN feedback, and in part also because of AGN feedback limiting their own growth. At $z=6$ the most massive MBH has $\mbh=1.3\times 10^6 \,\msun$. As a consequence, \Obelisk's AGN do not reach very high luminosity, the maximum luminosity being $3.5\times 10^{43} \, \rm{erg \, s^{-1}}$.

\begin{table*}
\caption{Models considered when studying properties of cLRDs.} \label{tab:models}
\centering
\begin{tabular}{c|p{0.22\linewidth}|p{0.22\linewidth}|p{0.22\linewidth}}
\hline \hline
Property &  &    &  \\
\hline
Column density & max: integrates the column density from the position of the MBH to the host galaxy virial radius  &  outmax80: when radiation pressure on dusty gas dominates excludes from the integration the inner 80~pc around the MBH   
&  \\
\hline
MBHs &  distr:  MBH masses are extracted from the \mbh-\mstar{} relation from G20 while Eddington ratios are extracted in log space from a normal distribution centered in $\log(0.25)$ with standard deviation 0.5 dex 
& distr overm: MBH masses are extracted from the \mbh-\mstar{} relation from P24 while Eddington ratios are extracted in log space from a normal distribution centered in $\log(0.25)$ with standard deviation 0.5 dex  & distr sEdd: Eddington ratios are extracted in log space  from a normal distribution centered in $\log(0.5)$ with standard deviation 0.5 dex  \\
\hline
dust & MW-like: a mass fraction of 54\% in silicates and 46\%
in carbonaceous grains with MRN size distribution cut out at $0.25\,\rm \mu m)$ &  SMC-like: 100\% silicates with MRN size distribution  &   \\
\hline
\end{tabular}
\tablefoot{ `max' integrates the column density from the position of the MBH to the host galaxy virial radius; `outmax80' integrates the column density from outside an 80-pc sphere around the MBH to the host galaxy virial radius to mimic the effect of radiation pressure on dust.}
\end{table*}

   \begin{figure}
   \centering
   \includegraphics[width=\columnwidth]{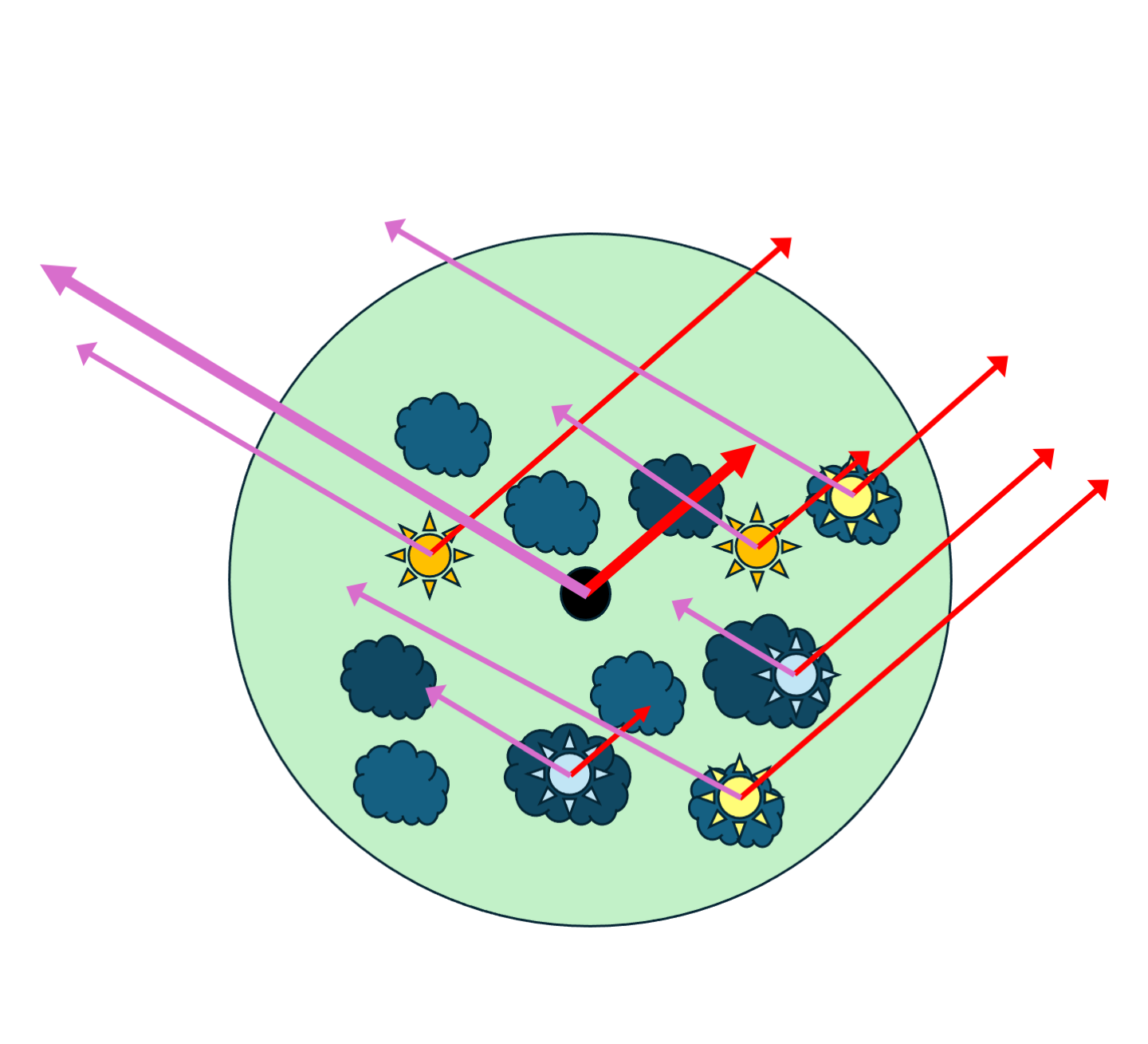}
   \caption{Sketch of our approach to modeling gas and dust absorption, showing how differential attenuation affects different parts of a given galaxy and lines of sight. We highlight two lines of sight with rays shown in purple and red arrows for each emitter. Darker grey colors show gas and dust clouds with larger optical depth. Stars' colors indicate age: for instance young (blue) stars are generally still embedded in their birth clouds, while this is not necessarily the case for older (yellow and orange) stars. Since at least some star-forming regions are dusty, their contribution to the SED will be more affected by the extinction curve than other parts of the galaxy. The position of the AGN is marked in black, with thicker arrows. The example distribution of emitters and absorbers shown here has been chosen to highlight how the same source can be AGN- or stellar-dominated, depending on the line of sight. In this example, along the cumulative purple line of sight the AGN dominates over starlight: the AGN is seen through a clear line of sight, while much of the emission from the stellar population is absorbed. Vice-versa, along the cumulative red line of sight the AGN emission is sub-dominant to starlight.}
              \label{fig:cartoon}
    \end{figure}

\subsection{Galaxy and AGN absorption and photometry}

A sketch of our approach is shown in Fig.~\ref{fig:cartoon}. For galaxies in \Obelisk{} besides intrinsic properties, e.g., stellar mass (\mstar), SFR, dust-to-gas ratio, photometry has also been calculated in NIRCAM bands (F115W, F150W, F200W, F277W, F356W, F444W), both considering the intrinsic emission determined by the star formation history and by attenuating it, as described in Trebitsch et al. (in prep). Stellar population SEDs are based on Bpass v2.2.1 binary stellar population model \citep{2017PASA...34...58E,2018MNRAS.479...75S}. They depend on the age and metallicity of the stellar population. The intrinsic SED is sampled using $10^6$ photons that probe the distribution of the luminosities of, and are assigned to, star particles, and each photon is assigned a restframe wavelength $\lambda$ in the range $0.075- 1 \, \mu\rm m$ based on the SED of its emitting star particle, as well as an isotropically selected initial direction. Photons are then propagated using the \textsc{Rascas} code \citep{MichelDansac2020}, based on the optical depth along the line of sight, computed using the dust mass density in each cell. In the case where a photon is absorbed, then the propagation stops, otherwise the photon is scattered in a new direction. Photons are collected in 12 independent directions to construct the observed SEDs.

A full discussion of galaxy dust properties and comparison with observations is presented in Trebitsch et al. (in prep). We here note only that the dust masses of $z=4-9$ in \Obelisk{} reproduce well the observed scaling and normalisation with stellar mass. In terms of dust attenuation, the typical $A_V$ increases from about $0.3$ to $1.5-2.0$ for galaxies with masses from $10^8{\rm M_\odot}$ to $10^{10}{\rm M_\odot}$ (depending on the choice of the extinction curve), with a dispersion of about $0.7$ mag around this value. Finally, the observed UV $\beta$ slopes (not accounting for nebular emission) range from $\beta \simeq -3$ to $-0.5$ assuming a SMC-like extinction curve, and to $-1.5$ for a MW-like extinction curve. We note that even for a given extinction curve, \Obelisk{} reproduces a variety of attenuation curves due to the complex geometrical effects between the stellar and dust distributions.

The AGN continuum is described by the following equation \citep{2017ApJ...849..155V}:
\begin{equation}
f_\nu = {\cal N}\left(\nu^{\alpha_{\rm UV}} e^{-\frac{h\nu}{kT_{\rm peak}}}e^{-\frac{kT_{\rm IR}}{h\nu}}+a \nu^{\alpha_{\rm X}}\right),
\label{AGN_SED}
\end{equation}
with $\nu$ the photon frequency, $h$ the Planck constant, $\alpha_{\rm UV}=0.5$, $\alpha_{\rm X}=1$, and $k T_{\rm IR}=0.01\,\rm Ryd$. The continuum is based in optical/near-IR on the Shakura-Sunyaev solution \citep{1973A&A....24..337S}, following \cite{2016ApJ...833..266T} for the estimation of $T_{\rm peak}$. $T_{\rm peak}$ represents the location of the peak of the spectrum, and it occurs at temperature of the order of $3\times 10^6-10^6-7\times 10^4$~K for black holes with $f_{\rm Edd}=0.1$ and mass $10^4-10^6-10^9$ respectively. As per the standard Shakura-Sunyaev solution, $T_{\rm peak}$ increases with decreasing black hole mass and increasing $f_{\rm Edd}$. The formalism in \cite{2016ApJ...833..266T} allows one to obtain a good match with standard bolometric corrections in the luminosity limit where the corrections have been derived. We refer the reader to Fig.~1 and Appendix A in \citet{2017ApJ...849..155V} for additional information. The model is calibrated in X-rays using results from the physical models developed by \cite{2012MNRAS.420.1848D}: the normalization $a$ is obtained through $\alpha_{\rm OX}$, the exponent of a power-law connecting the continuum between 2 keV and $2500\, \AA$, fitting the dependence on MBH mass and Eddington ratio using the results in \cite{Dong2012}.  The last term in Equation~\ref{AGN_SED} describes the corona contribution and it is set to zero below $0.1 T_{\rm peak}$, where the contribution becomes negligible: this avoids that the power-law pops up at low frequencies below the peak at $T_{\rm peak}$. The global normalization ${\cal N}$ is obtained by requiring that the bolometric luminosity matches the accretion luminosity 
$L_{\rm bol}=1.26\times 10^{38} f_{\rm Edd} (M_{\rm BH}/{\rm M_\odot}) \, \rm{erg \, s^{-1}}$.
X-ray luminosity is calculated in the $[2-10]\,\rm keV$ range (observer frame). We have not attempted to include scattered light contributing to the UV emission. 

We consider two extinction laws: one similar to the Milky Way (`MW-like') with $54\%$ of the grain number density in silicate grains and $46\%$ in carbonaceous grains, and one similar to the Small Magellanic Cloud (`SMC-like') with $100\%$ silicate grains. They both use the \cite{Mathis77} grain size distribution ranging from $0.001 \,\rm \mu m$ to $0.25\,\rm \mu m$ (slope -3.5), and dust grain optical constants from \cite{2001ApJ...548..296W} and \cite{1993ApJ...402..441L}. For a given amount of dust, the MW-like extinction curve, compared to the SMC-like, leads to more extinction overall but to much less reddening (i.e. to less extinction in the blue part, $\lambda \lesssim 0.2\,\rm \mu m$ restframe, relative to the red part of the spectrum, $\lambda \gtrsim 0.25\,\rm \mu m$ restframe). To model absorption for AGN we calculate the contribution of the interstellar medium (ISM) to the column density at the location of the MBHs by casting rays isotropically around each MBH in the outputs of the simulation.  We used a version of the \textsc{Rascas} code \citep{MichelDansac2020} modified to integrate column densities very efficiently.  Values of 12 individual rays per galaxy, for both the stellar population and the AGN, are employed throughout the rest of the paper unless stated otherwise. To estimate the AGN attenuation at JWST wavelengths we multiply the hydrogen column density by the mean dust-to-gas mass ratio within $2R_{\rm 50}$, where $R_{\rm 50}$ is the 3D half-mass radius, to obtain the dust column density and then convolve it with the extinction laws to obtain the optical depth. We have checked that there is no significant gradient in the dust-to-gas mass ratio between $R_{\rm 50}$ and $2R_{\rm 50}$.  The typical dust-to-gas mass ratios are between $10^{-2}$ and $10^{-3}$. 

To estimate the AGN gas column density we first integrated the gas column density from the MBH position to the virial radius of its host halo for each of the 12 rays (`max' column density).  This results in very high gas column densities, making the AGN contribution in JWST bands completely negligible and producing a very high fraction of obscured AGN in X-rays. Indeed, the gas distribution near the MBHs is stochastic  -- depends on the position of few gas cells -- and the inclusion of boosted gas dynamical friction makes the MBH cling to dense gas. Furthermore, simulations find that the column density around an AGN decreases after feedback has been able to clear out paths in its surroundings \citep{2019MNRAS.487..819T, 2020MNRAS.495.2135N,2022MNRAS.510.5760L}. In \Obelisk{} MBHs start to grow when galaxy masses reach $\mstar\sim 10^9 \, \msun$, this is mainly because SN feedback makes gas very turbulent in low-mass galaxies. This retarded growth gave a better match to pre-JWST AGN luminosity functions \citep[][see also Fig.~\ref{fig:bolLF}]{Trebitsch2021}, but as MBHs struggle to grow they have not been able to open these chimneys in most of \Obelisk's galaxies. To account for these considerations, we base our fiducial `outmax80' model on the proposal of \citet{2008MNRAS.385L..43F} that the radiative pressure on dust evacuates dusty gas in the MBH vicinity at high Eddington ratios, or that it prevents its accumulation \citep{2009MNRAS.394L..89F}. \cite{2008MNRAS.385L..43F} show that for a given Eddington ratio there is a minimum  column density for gas to counteract radiation pressure and not be blown away. In practice, this provides a limiting column density value as a function of the Eddington ratio below which dusty gas will be removed or prevented to accumulate. We include this effect in the model as follows. We fit the curve of this threshold column density as a function Eddington ratio \citep{2008MNRAS.385L..43F} and compare it to the column density seen by the MBH. If the `max' column density is below this threshold value, we consider that the inner region has been evacuated and integrate the column density outside a sphere of radius\footnote{We have also tested spheres with radius $40\,\rm pc$ and $140\,\rm pc$, and selected $80\,\rm pc$ because it gave the best match to the absorbed fraction as probed by X-ray observations, see Section~\ref{sec:SED_abs}.} $80 \,\rm pc$. Otherwise we retain the full `max' column density. In the following we consider this as the fiducial case for the column density, since it captures the effect that feedback from actively growing MBHs would have on the galaxy, effect that is not present in \Obelisk's galaxies because of the limited MBH growth.

 \begin{figure*}
   \centering
   \includegraphics[width=\columnwidth]{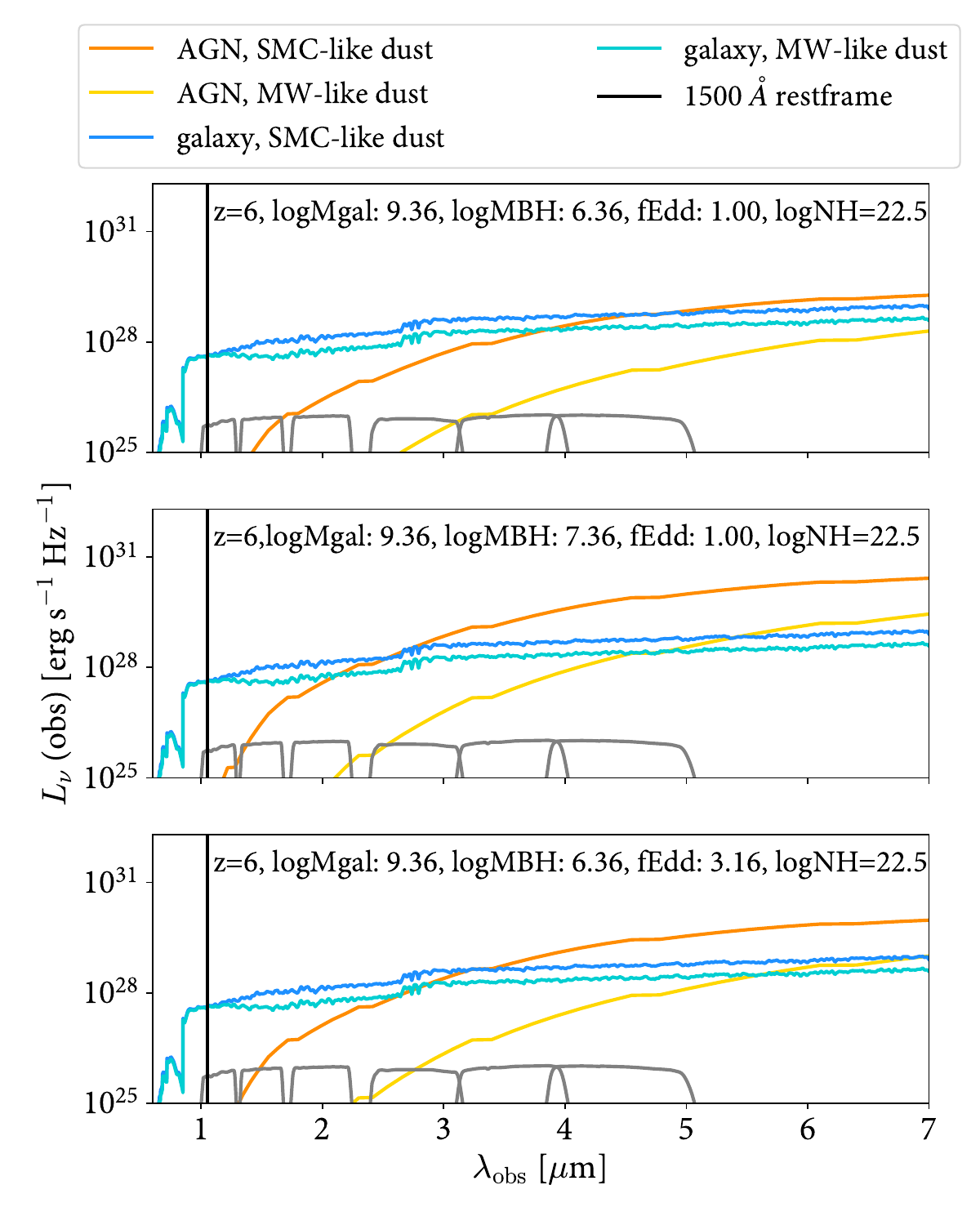}
   \includegraphics[width=\columnwidth]{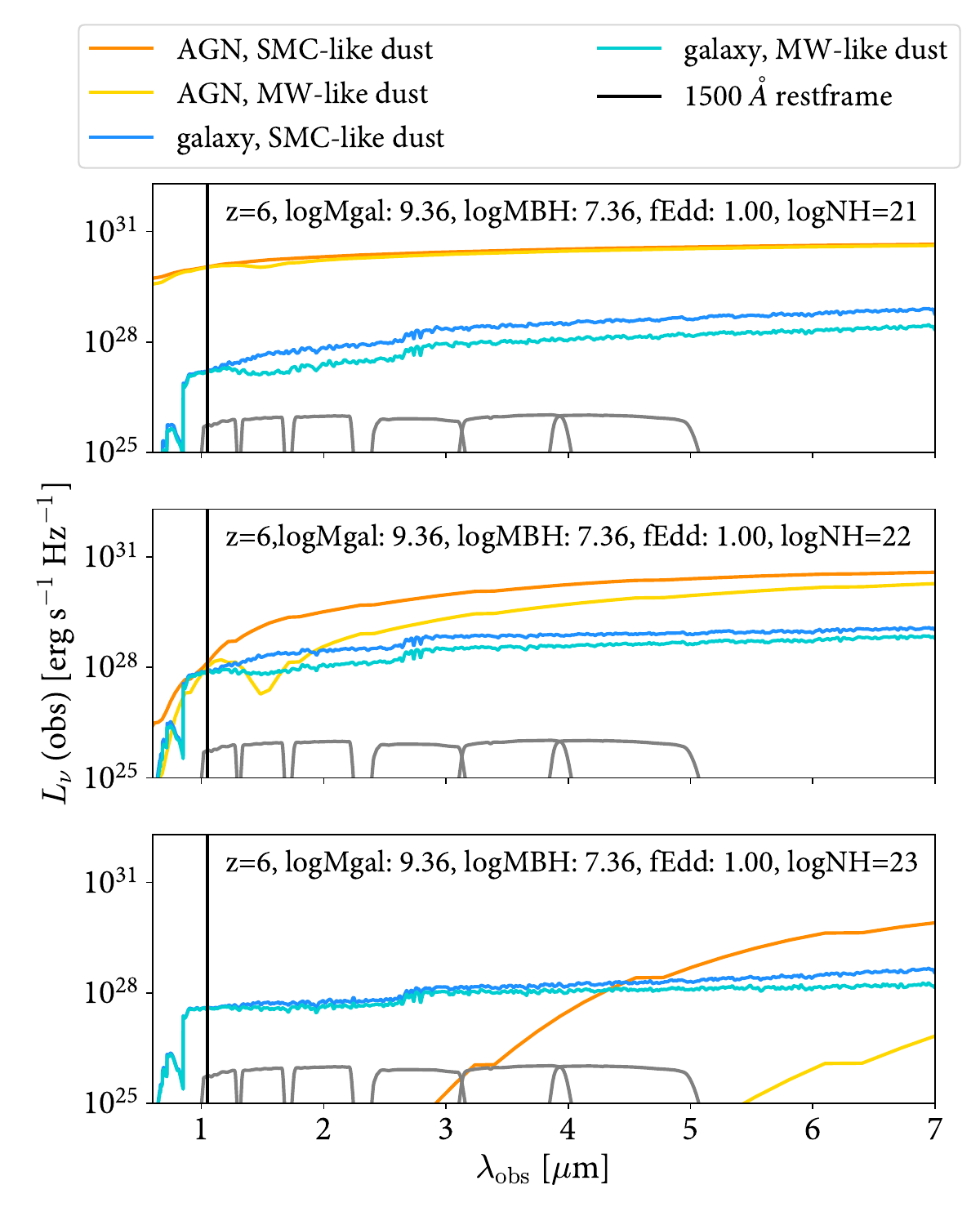}
     \caption{Examples showing how different assumptions affect the emergent SEDs at $z=6$.  For the same galaxy we show different \mbh, \fedd{} and $N_{\rm H}$, as well as different lines-of-sight for the emission from the stellar population. The NIRCAM response for the filters analyzed in this paper is shown in gray.}
              \label{fig:SEDs}
    \end{figure*}

To estimate the X-ray obscuration properties we use the hydrogen column density, and we furthermore consider that dust-free gas can linger between the edge of the accretion disc, calculated as the self-gravity radius, and the dust sublimation radius. As noted by \citet{2024arXiv240500504M} the mismatch between high hydrogen column densities and low dust column densities is a typical feature even in low-redshift AGN, and the addition of this gas eases the tension between the X-ray and optical AGN populations. However, one has to justify that just outside the dust sublimation radius radiation does not find a wall of dusty gas, which would happen if the density distribution were continuous.  The `outmax80' model we adopt, where radiative pressure on dust evacuates dusty gas, would explain how a very high gas density within the dust sublimation radius can drop to low dusty gas densities immediately afterwards.  To include the unresolved contribution below the dust sublimation radius we follow the approach of \citet{2023A&A...676A...2D}. Assuming that gas falls radially under the effect of the MBH gravity and a spherically symmetric configuration the gas density in the torus should follow a power-law density profile $\rho\propto r^{-3/2}$. The density profile is normalized so that the total mass between the Bondi radius and the edge of the accretion disc, which we assume corresponds to the self-gravity radius, corresponds to the accretion rate multiplied by the free fall time. For the self-gravity radius we adopt the expression from \citep{Laor1989}:

\begin{equation}
    r_\mathrm{sg} \simeq 3586 \, r_{\rm g} \alpha^{2/9}\left(\frac{\mbh{}}{10^8 {\rm M_\odot}}\right)^{-2/9} \fedd^{4/9}\, ,
    \label{eq:rsg}
\end{equation}
where $r_{\rm g}=G \mbh/c^2$, $G$ is the gravitational constant, $c$ is the speed of light in vacuum, and we set $\alpha=0.1$. We then integrate the density profile from $r_\mathrm{sg}$ to the dust sublimation radius \citep{Suganuma06}: 
\begin{equation}
    r_\mathrm{subl} \approx 0.47 \left(\frac{L_\mathrm{UV}}{10^{46}\rm erg\, s^{-1}}\right)^{1/2} \,  \si{\parsec}. 
\end{equation}
The final expression is:
\begin{equation}
\label{eq:nh_subl}
   N_{\rm H, inner}=\frac{3 \sqrt{2}\pi}{4 \varepsilon_r \, c\, \sigma_{\rm T}} \fedd \, (G \mbh)^{1/2} \,r_\mathrm{subl}^{-1/2}\left[\left(\frac{r_\mathrm{sg}}{r_\mathrm{subl}}\right)^{-1/2}-1\right],
\end{equation}
where $\varepsilon_r$ is the radiative efficiency and $\sigma_{\rm T}$ is the Thomson cross section. For ease of implementation we fit $N_{\rm H, inner}$ from the sample of MBHs in galaxies with mass $\mstar>10^8\, \msun$ and $\fedd>10^{-4}$ at $z=6$ to obtain:
\begin{equation}
\label{eq:nhvsfeddmstarspin}
    \log\left(\frac{N_{\rm H, inner}}{\rm cm^{-2}}\right)=0.72\log(\fedd)-0.29\log\left(\frac{\mbh}{10^8{\rm M_\odot}}\right)+24.90
\end{equation}
with RMS residual = 0.16 when we consider the actual value of $\varepsilon_r$ from \Obelisk{}, which includes spin evolution, and in which spin increases with MBH mass (so radiative efficiency decreases), as discussed in \citet{2023A&A...673A.120D}. If we consider a fixed $\varepsilon_r=0.1$, we find 
\begin{equation}
\label{eq:nhvsfeddmstar}
    \log\left(\frac{N_{\rm H, inner}}{\rm cm^{-2}}\right)=0.77\log(\fedd)+0.10\log\left(\frac{\mbh}{10^8{\rm M_\odot}}\right)+23.13
\end{equation}
with RMS residual = 0.001, but in this case all AGN with $L_{\rm 2-10keV}>3\times 10^{43} \,{\rm erg\, s^{-1}}$ would have $ N_{\rm H, inner}$ above $10^{23}$~cm$^2$. We note, however, that our model of this dust-free gas is very simplistic, assuming spherical symmetry and a density power-law slope of $-3/2$, therefore we consider this more of an indication of the possible magnitude of the effect than an actual quantitative estimate. For the remainder of the paper we consider Eq.~\ref{eq:nhvsfeddmstarspin} when we include unresolved obscuration.
   
    \begin{figure}
   \centering
   \includegraphics[width=\columnwidth]{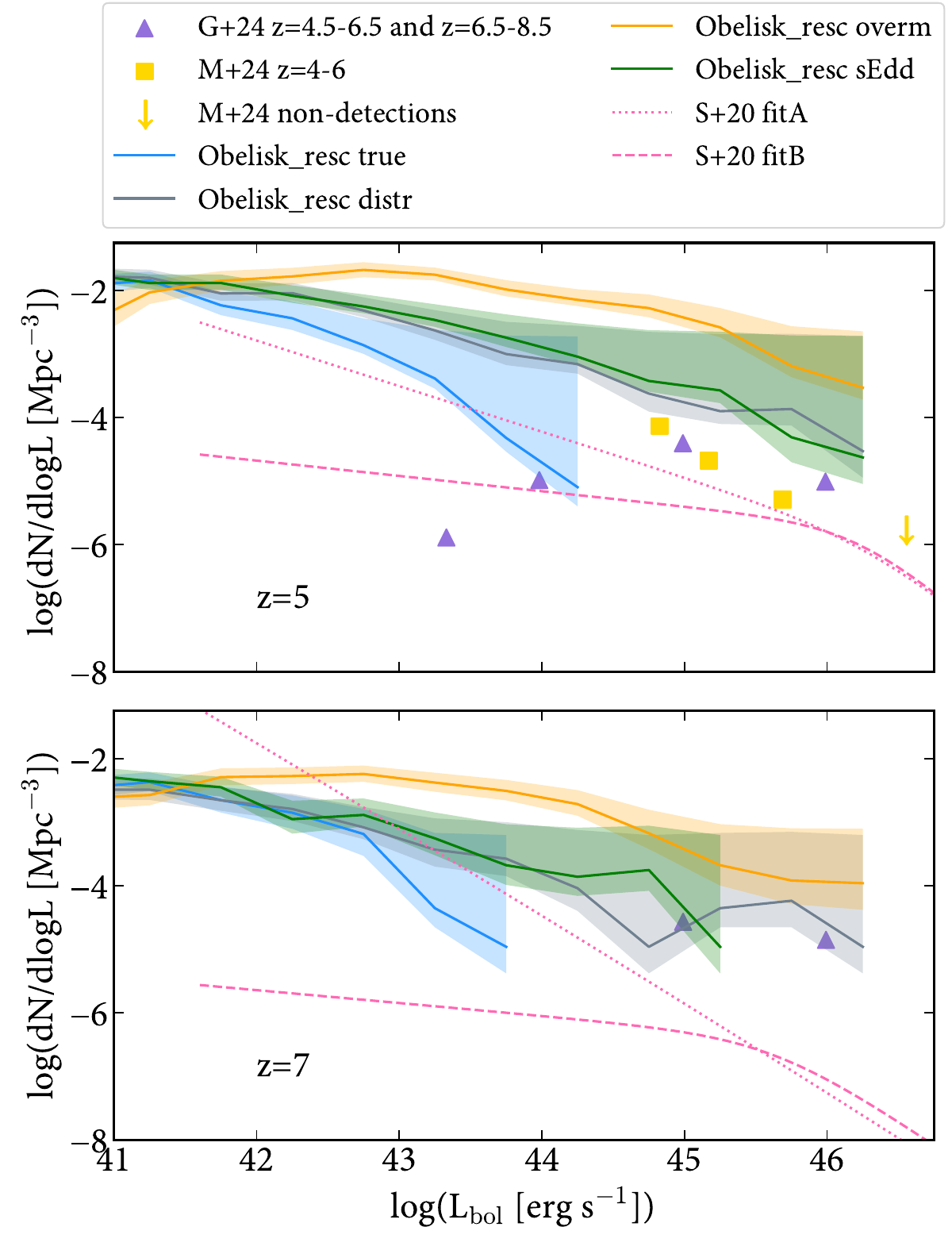}
   \caption{Bolometric luminosity function for the actual MBHs in the simulation (`true') and for the models described in  Table~\ref{tab:models}, where `resc' means we have corrected the number densities for the overdensity bias: \Obelisk{} is a protocluster. To obtain the cosmic average we have rescaled the \Obelisk{} galaxy mass function to that of the \textsc{NewHorizon} simulation, which simulates an average region of the Universe. This figure assumes an active fraction of unity, and therefore represents an upper limit to the LF for each of the models. The results are compared to the luminosity functions derived in \citet[][M+24]{2024ApJ...963..129M} and \citet[][G+24]{2024ApJ...964...39G} from JWST data and to fits to the evolution of the bolometric luminosity function derived in \citet[][S+20, pre-JWST]{2020MNRAS.495.3252S}. }
              \label{fig:bolLF}
    \end{figure}
  
\subsection{Massive black hole population}
 The `true' MBH properties in the simulation do not contain any overmassive MBHs that can be compared to JWST AGN candidates, and many MBHs accrete at low Eddington ratios, therefore we create a population, `distr' where the MBH masses are extracted from the \mbh-\mstar{} relation from \citet[][G20]{2020ARA&A..58..257G}:
 \begin{equation}
\log\left(\frac{\mbh}{\rm M_\odot}\right)=6.87+1.39 \log\left(\frac{\mstar}{10^{10}\rm M_\odot}\right).
\end{equation}
 We also consider an intrinsically overmassive population ('distr overm') based on the fit by \citet[][P23]{2023ApJ...957L...3P}:
\begin{equation}
\log\left(\frac{\mbh}{\rm M_\odot}\right)=8.17+1.06\log\left(\frac{\mstar}{10^{10}\rm M_\odot}\right).    
\end{equation} 
 By mocking the MBH population to obtain sufficiently luminous AGN we can address what type of MBH properties are needed to have cLRDs, given a normal population of galaxies. We cannot, however, address how cLRDs are born: that would require self-consistent modeling linking MBH and galaxy evolution, in particular the star formation history and how AGN emission and outflows modulate the dust properties \citep[e.g.,][]{2002ApJ...567L.107E,2004ApJ...616..147G,2023Galax..11..102A}. In practice, all properties related to galaxies and the interstellar medium are self-consistent, while for the AGN we retain only its position in the galaxy, which determines the gas and dust column density seen by the mocked AGN.
 
Based on \citet{2017ApJ...849..155V} and \citet{2024ApJ...964...39G}, Eddington ratios are extracted from a normal distribution centered in $\log(0.25)$ with standard deviation 0.5 dex for `distr' and 'distr overm', and centered in $\log(0.5)$ for a `sEdd' case, where if $\fedd>0.3$ we shift redwards the energy peak of the spectrum to mimic photon trapping, using $r_{\rm cor}=20$ for sub-Eddington and $r_{\rm cor}=200$ for super-Eddington \citep[see][for a fully physical model]{2019MNRAS.489..524K}. Here $r_{\rm cor}$ is in units of the MBH gravitational radius, and sets the peak temperature of the SED. Its value in the  sub-Eddington case has been calibrated to match standard bolometric corrections in the relevant limits \citep{2017ApJ...849..155V}. In either case we do not cap \fedd{} and we do not decrease the overall luminosity for sources with $\fedd>0.3$, but we have checked that limiting the luminosity to $L_{\rm Edd}$ does not change quantitatively our results.  The effect that quantitatively changes the results in the sEdd case is related to the relative increase in luminosity of the AGN with respect to the galaxy at the redder JWST bands (356, 444) with respect to the bluer ones (115, 200) because the energy peak in the sEdd case shifts toward lower frequencies.  

Model assumptions are summarized in Table~\ref{tab:models}. Since this analysis is mainly focused on AGN with bolometric luminosity $>10^{44} \,{\rm erg\, s^{-1}}$ we consider galaxies in the \Obelisk{} simulation at redshift $z=6$ and $\mstar>10^8 \msun$. Although we post-process \mbh{} and \fedd{} we  analyze galaxies hosting MBHs in the simulation, unless specified otherwise. The reason is that we need to know the MBH position in order to calculate attenuation. For the analysis of the absorbed fraction we extend the analysis to $z=4$ and $z=5$. For the bolometric luminosity function, we consider all galaxies with $\mstar>10^8 \msun$ at $z=5$ and $z=7$. In future work we will explore the redshift evolution of AGN and cLRDs.

Finally, at $z=6$ the simulation includes 340, 75 and 8 galaxies with $\mstar=10^8-10^9 \msun$, $10^9-10^{10} \msun$ and $>10^{10} \msun$ respectively. This means that the high-mass end of the sample is sparsely populated, and the results will be highly dependent on the randomly picked \mbh{} and \fedd{}  for the mock sample: drawing a single \mbh{} and \fedd{} per galaxy would not sample the full parameter space. We therefore draw a different \mbh{} and \fedd{} from the respective distributions for each of the 12 lines of sight in order to obtain a sufficiently large sample, except in Section~\ref{sec:los}, where we explore the effect of line-of-sight differences.  At $z=4$ and $z=5$, used only to estimate the X-ray absorbed fraction, we draw 2 and 4 cases per galaxy, using the mean column density rather than 12 lines of sight, in order to have a similar sample size at each redshift.  

To validate the model and show that it produces a reasonable AGN population, the AGN bolometric luminosity functions at $z=5$ and $z=7$ (for $\mstar>10^6 \msun$ to avoid incompleteness) are shown in Fig.~\ref{fig:bolLF}. To obtain the cosmic average we have rescaled the \Obelisk{} galaxy mass function, which simulated an overdensity, to that of the \textsc{NewHorizon} simulation, which simulated an average region of the Universe. The rescaling of the mass functions provides a correction for the number density of galaxies as a function of their mass and redshift. To derive the rescaled AGN LF, we calculate the individual AGN LFs in different \mstar{} bins, renormalize the number density in each mass bin based on the correction calculated from the ratio of the \textsc{NewHorizon} and  \Obelisk{} galaxy mass function in that mass bin, and then sum the individual AGN LFs obtained in the different \mstar{} bins to obtain the complete LF. We have checked that we obtain similar results rescaling by the halo mass function, and binning in halo mass, but the number density correction has a steeper dependence on halo mass and therefore we opted for rescaling based on galaxy mass to have a smoother evolution.

The  simulated AGN LF  is compared in Fig.~\ref{fig:bolLF} to the bolometric luminosity function of AGN from \citet[][M+24]{2024ApJ...963..129M},  
\citet[][G+24]{2024ApJ...964...39G} and \citet[][S+20]{2020MNRAS.495.3252S}. The observational estimate of the luminosity function (LF) around becomes more uncertain as the luminosity decreases. 
The spectroscopic analysis in FRESCO and EIGER in M+24 did not detect any broad line AGN with bolometric luminosity larger than $\sim 10^{46}\, {\rm erg\, s^{-1}}$, therefore this gives an upper limit at high luminosity (yellow downward arrows in Fig.~\ref{fig:bolLF}). On the other hand, these bolometric LFs have been derived from broad line AGN only, and \citet{2023arXiv231118731S} finds a comparable-size population of narrow-line AGN. The extrapolations of some of the LFs would likely overpredict the bright end of the LF as known pre-JWST. Note, however, that for ease of interpretation in Fig.~\ref{fig:bolLF} and throughout the paper we have assumed an active fraction of unity: each galaxy hosts an MBH that is accreting at a non-negligible Eddington ratio. This is therefore an upper limit. Assuming that only 10\% of MBHs independently of MBH or galaxy mass are active, for instance, would decrease the LF by a factor of 10. We discuss how an active fraction of 100\% translates into different detectable AGN fractions in Section~\ref{sec:props}. 

   \begin{figure}
   \centering
  \includegraphics[width=\columnwidth]{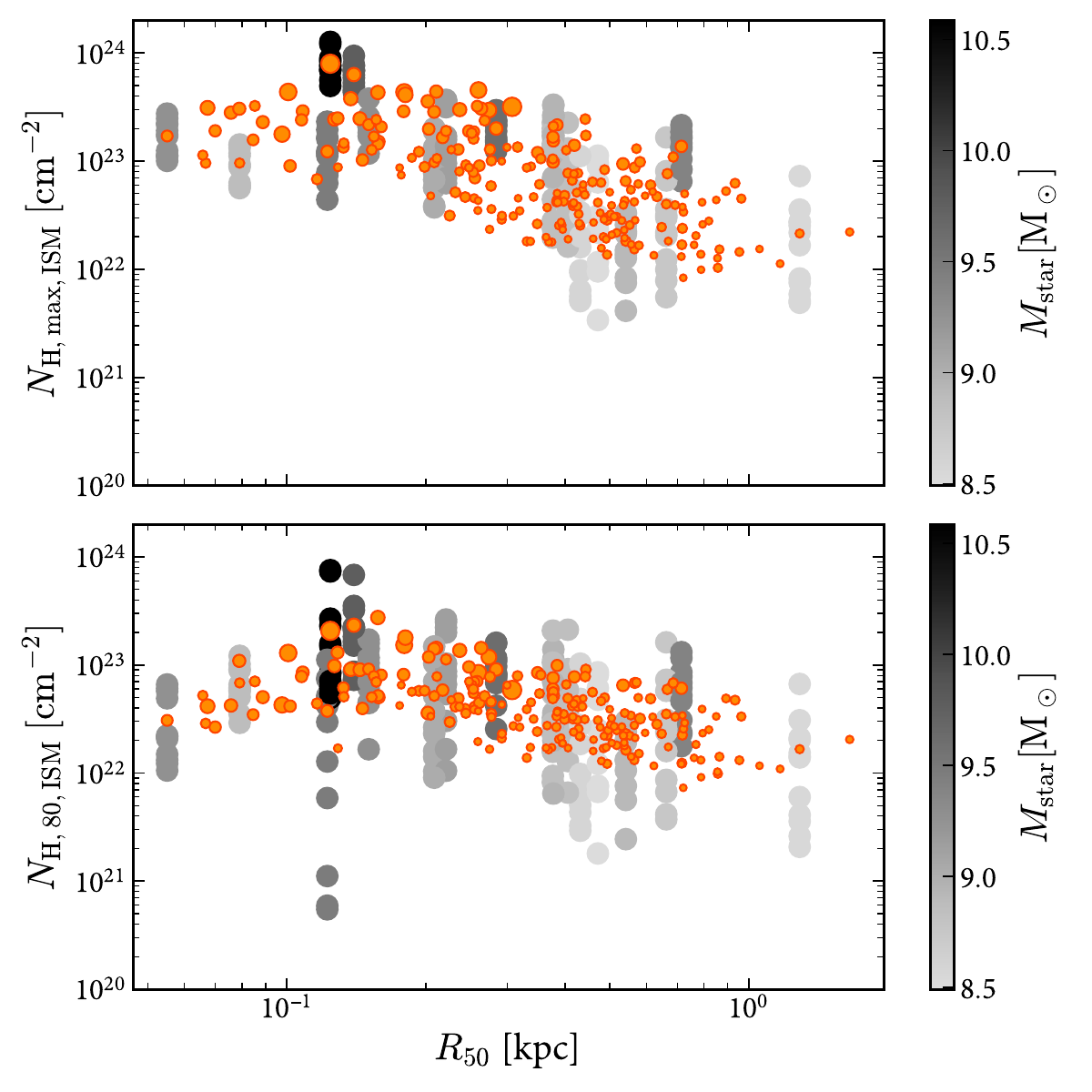}
   \caption{Relation between hydrogen column density in the ISM and galaxy mass and half-mass radius at $z=6$. Top panel:  column density obtained 
integrating the column density from the position of the MBH to the host galaxy virial radius (`max' model). Bottom panel: column density obtained excluding the inner 80~pc around the MBH (`outmax80' model). The orange points show the mean column density for each galaxy, with size scaling with \mstar. For a random subset of 24 galaxies we show the column density for 12 different lines of sight, with the greyscale proportional to \mstar, as shown in the colorbar.}
              \label{fig:NH_size}
    \end{figure}

\begin{figure}
   \centering
   \includegraphics[width=\columnwidth]{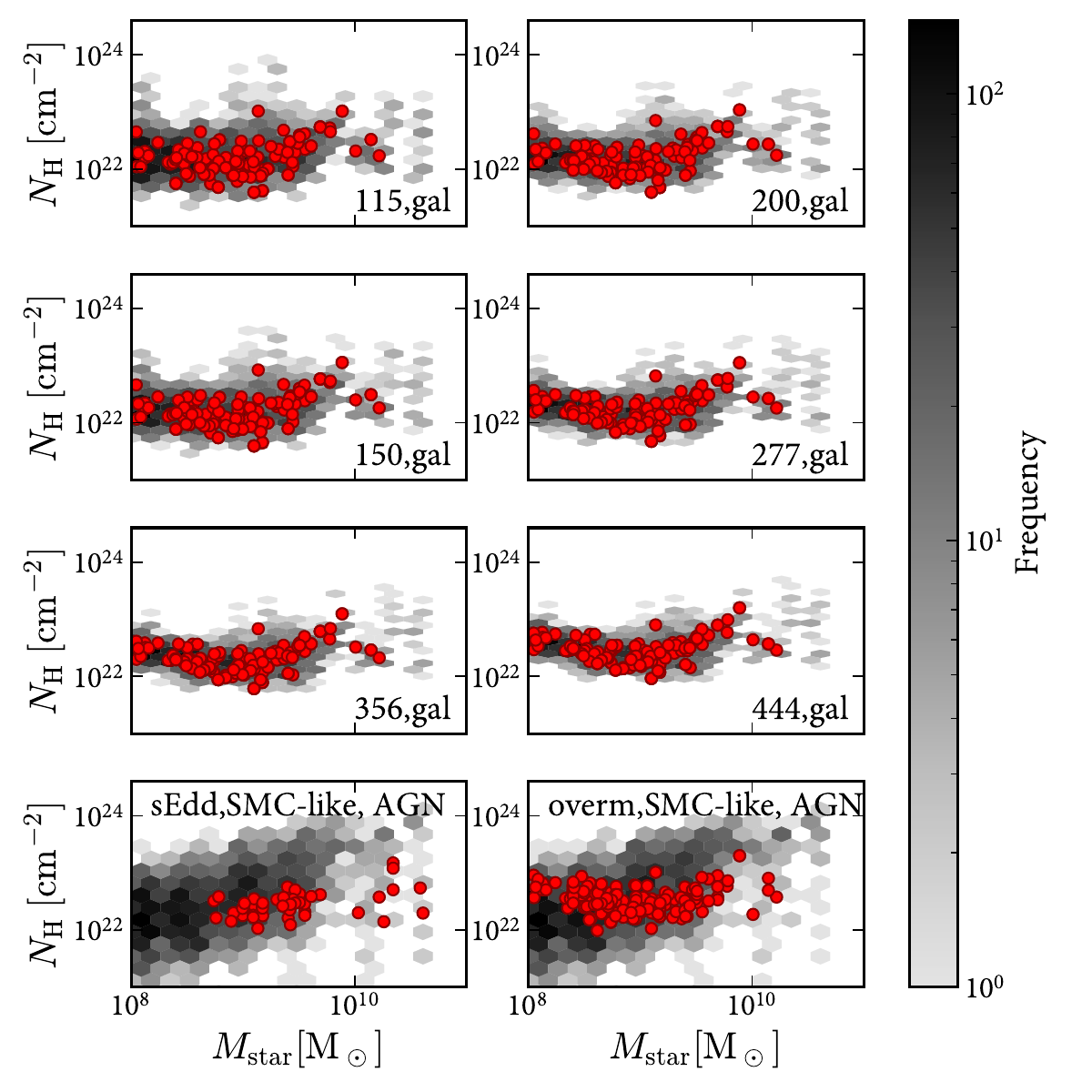}
   \caption{Top three rows: distribution of `effective' hydrogen column density for starlight obtained from the ratio of intrinsic and attenuated fluxes in each JWST filter we have considered (Eq.~\ref{eq:effNH}). Bottom row: distribution of hydrogen column density seen by the AGN as a function of galaxy stellar mass at $z=6$. This is the column density integrated above the dust sublimation radius, i.e. the column density that enters into calculating dust absorption.  The red points highlight the cases that host a cLRD meeting the UNCOVER selection.}
              \label{fig:NHdistr}
    \end{figure}

      \begin{figure}
   \centering
   \includegraphics[width=\columnwidth]{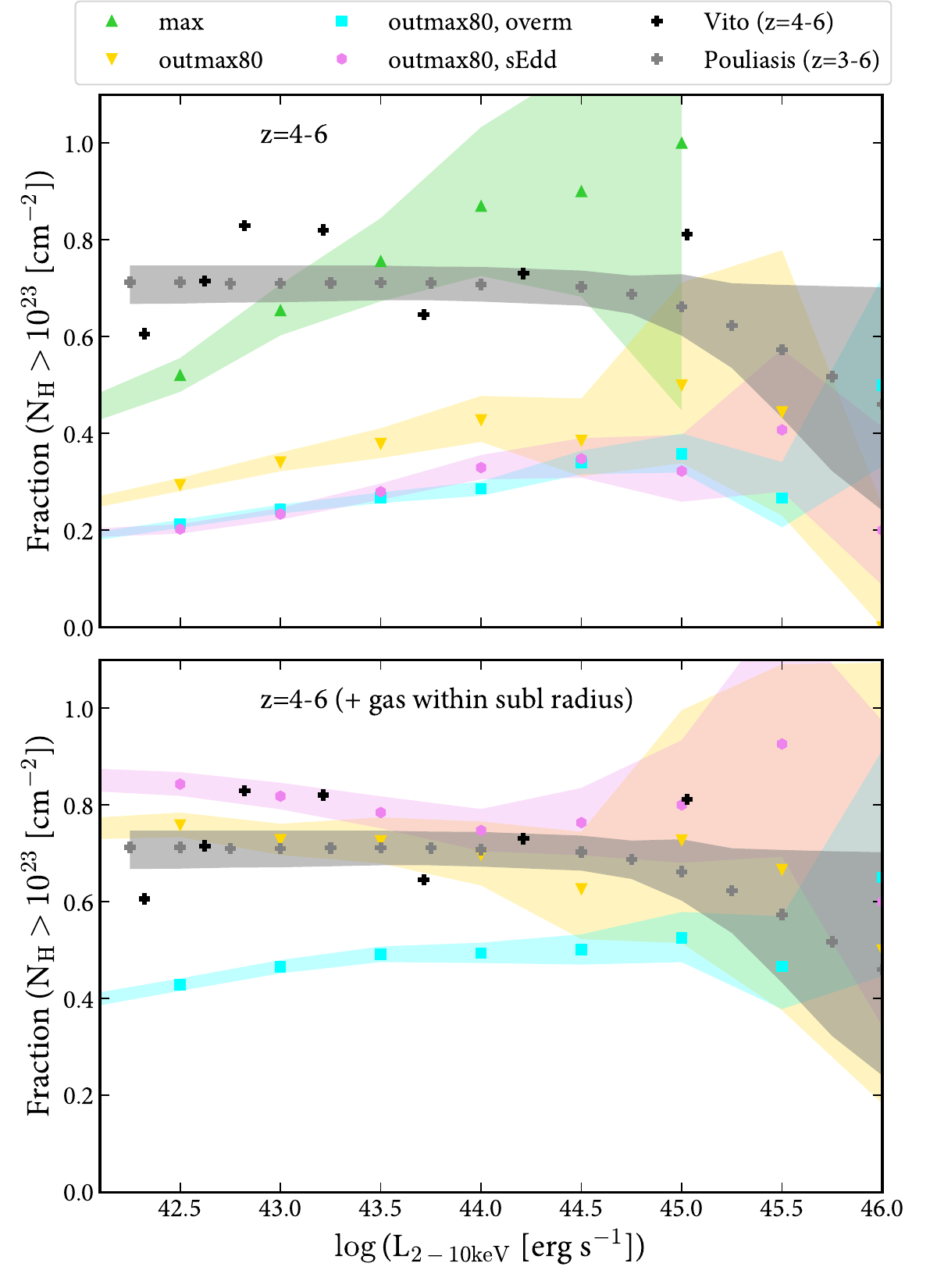}
   \caption{Absorbed fraction, defined as the fraction of AGN with ${\rm NH}>10^{23} {\rm cm}^{-2}$  for various models compared to \citet{2018MNRAS.473.2378V} and \citet{2024A&A...685A..97P}. We include the 68\% confidence region as shaded areas. The top panel shows the results including only the ISM gas, the bottom panel adds gas within the dust sublimation radius. For the cases including gas within the dust sublimation radius, we show the case with $\varepsilon_r=0.1$, including spin evolution the absorbed fraction decreases to 60-70\% (distr and sEdd), and 40\% (overm).}
              \label{fig:absfrac}
    \end{figure}

  \begin{figure}
   \centering
   \includegraphics[width=\columnwidth]{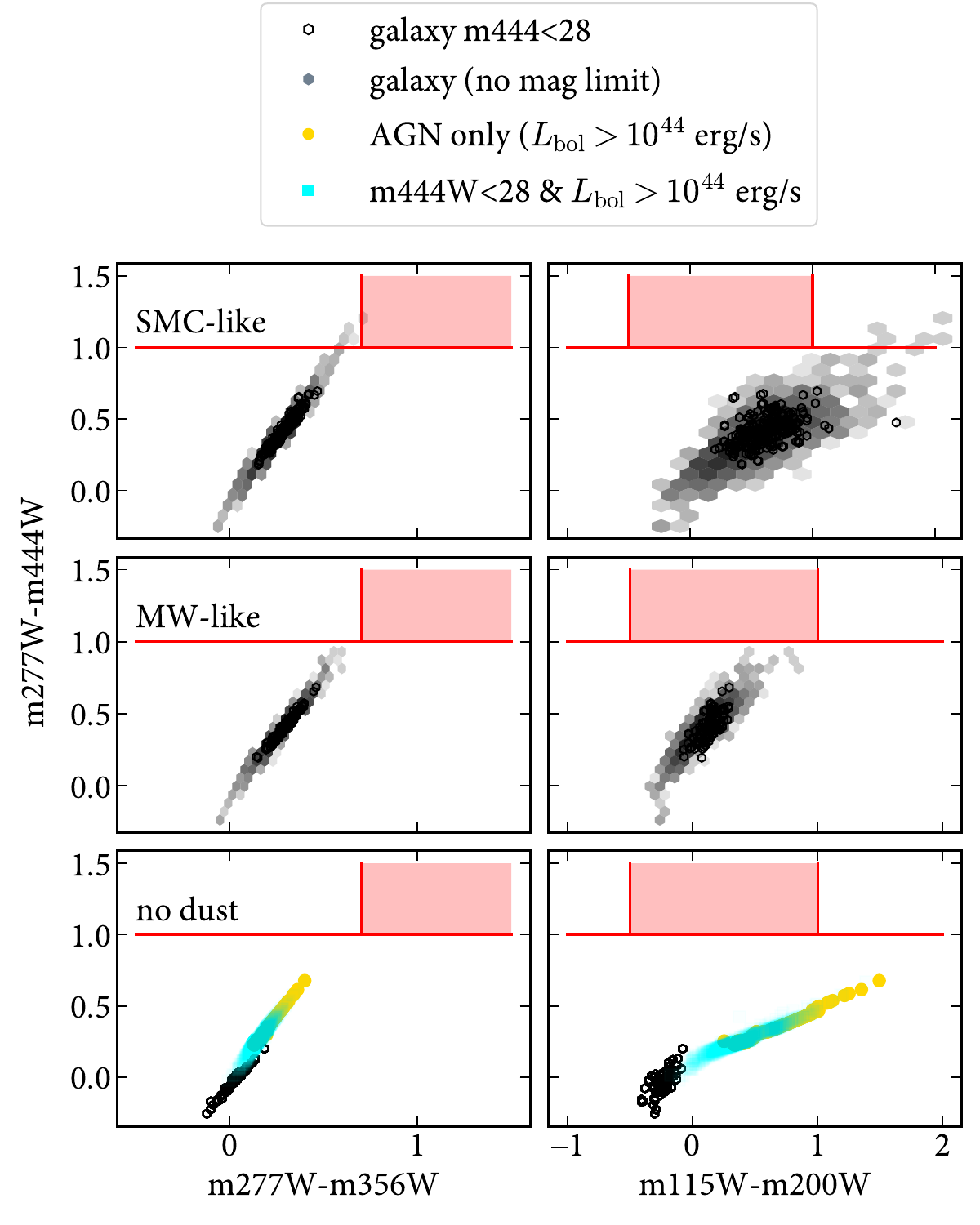}
   \caption{Colors for galaxies only for SMC- (top) and MW-like  (middle) dust extinction curves, and for galaxies and AGN without including dust (bottom). Galaxies at $z=6$ do not enter the UNCOVER color selection highlighted by the red shaded area, nor do AGN without inclusion of dust. We consider here the `overm' case, which leads to the largest AGN contribution to colors.}
              \label{fig:colors_dust_nodust}
    \end{figure}

   \begin{figure*}
   \centering
   \includegraphics[width=0.9\textwidth]{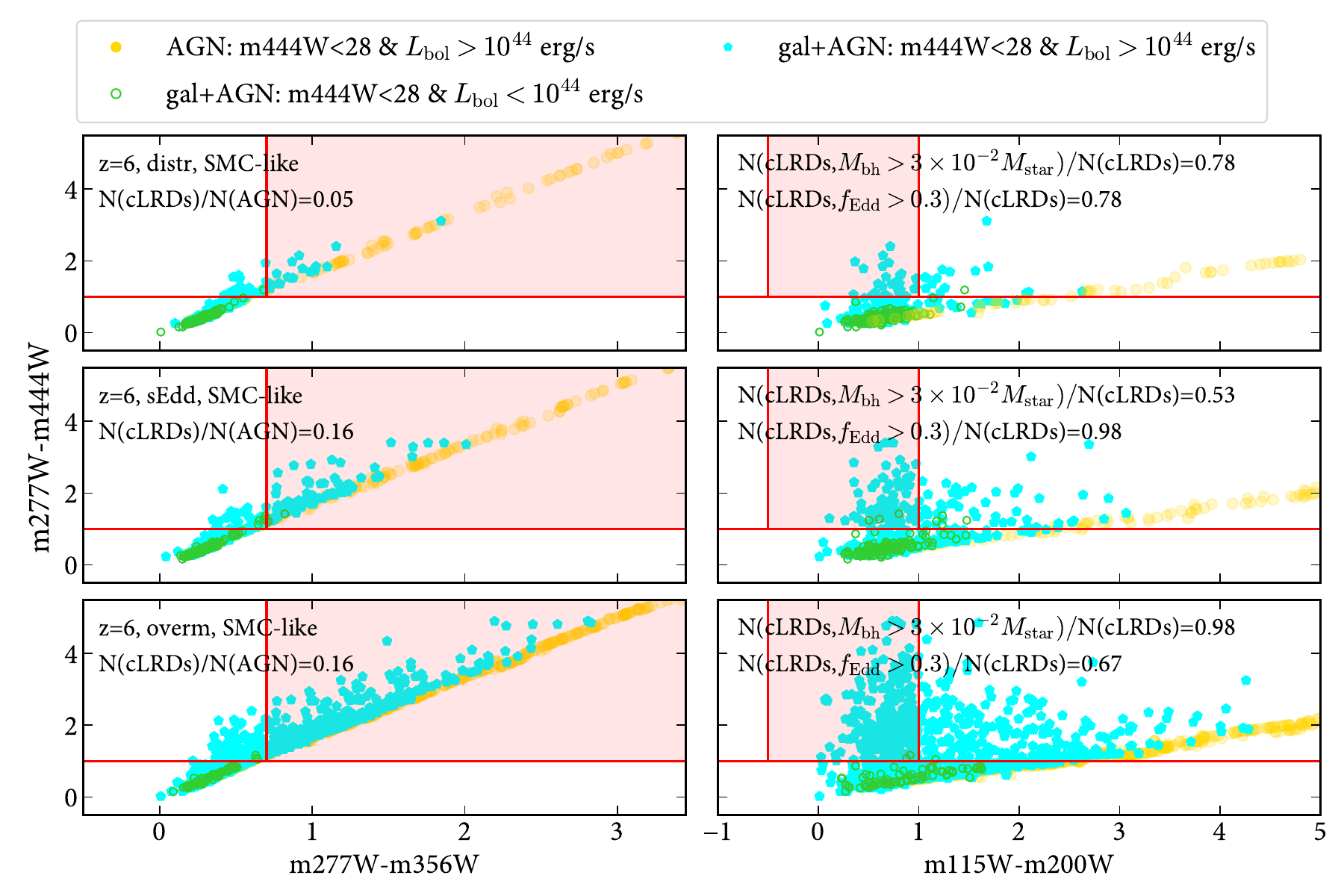}
 \caption{Colors for SMC-like dust. AGN are defined as sources having $L_{\rm bol}>10^{44} \, {\rm erg \, s}^{-1}$ and m444W<28 (combined galaxy+AGN). cLRDs are defined as the AGN that fall in the UNCOVER color selection (red shaded area). In the left panels we report the fraction of cLRDs over AGN.  The expected fraction is between 10 and 20\%. In the right panels we report the fraction of cLRDs with \mbh$> 3\times 10^{-2}$\mstar) and with \fedd$>0.3$. The yellow points report the colors considering only the AGN emission. }
              \label{fig:colors_z6_SMC}
    \end{figure*}
 
   \begin{figure*}
   \centering
   \includegraphics[width=0.42\textwidth]{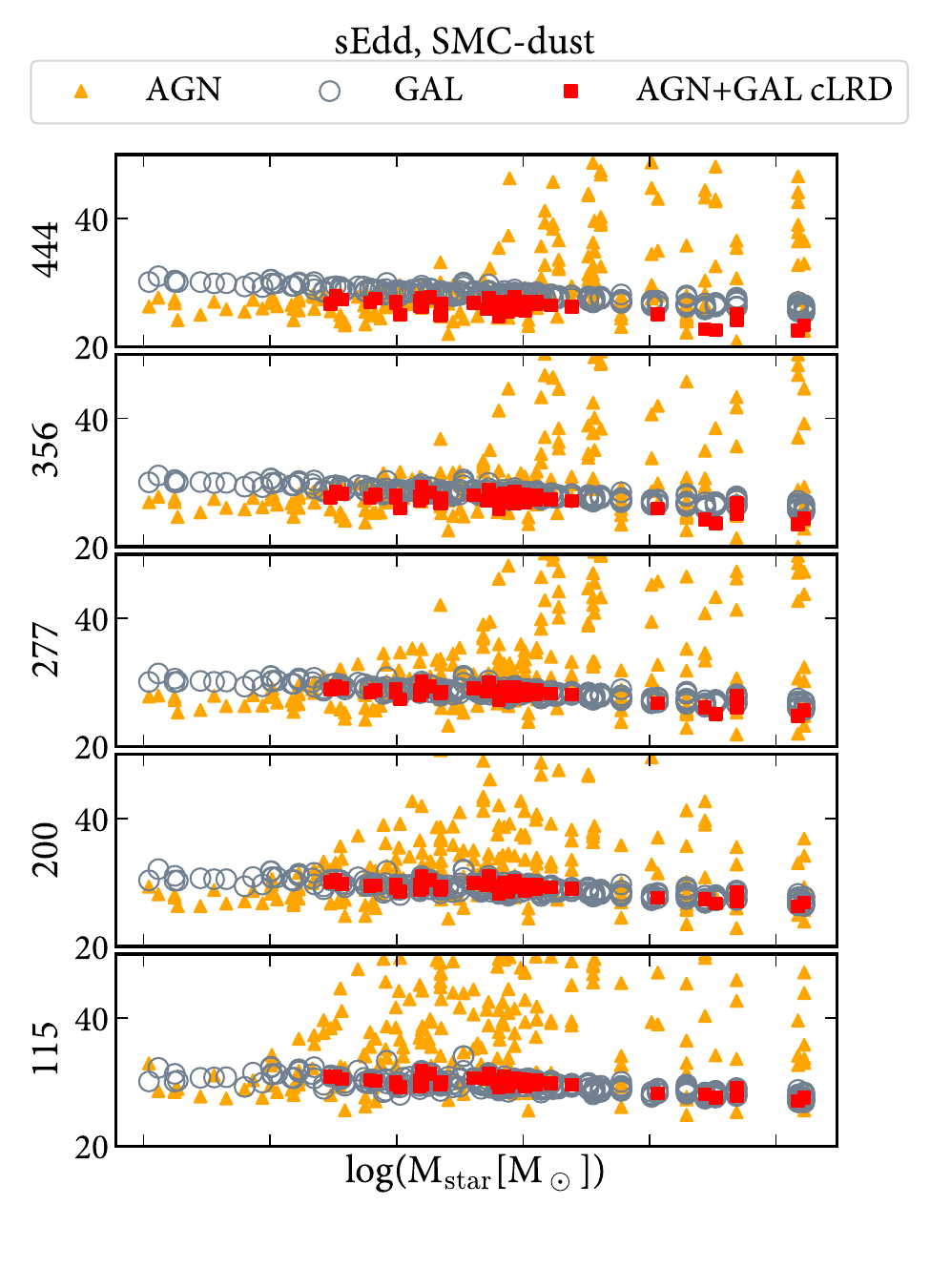}
   \includegraphics[width=0.42\textwidth]{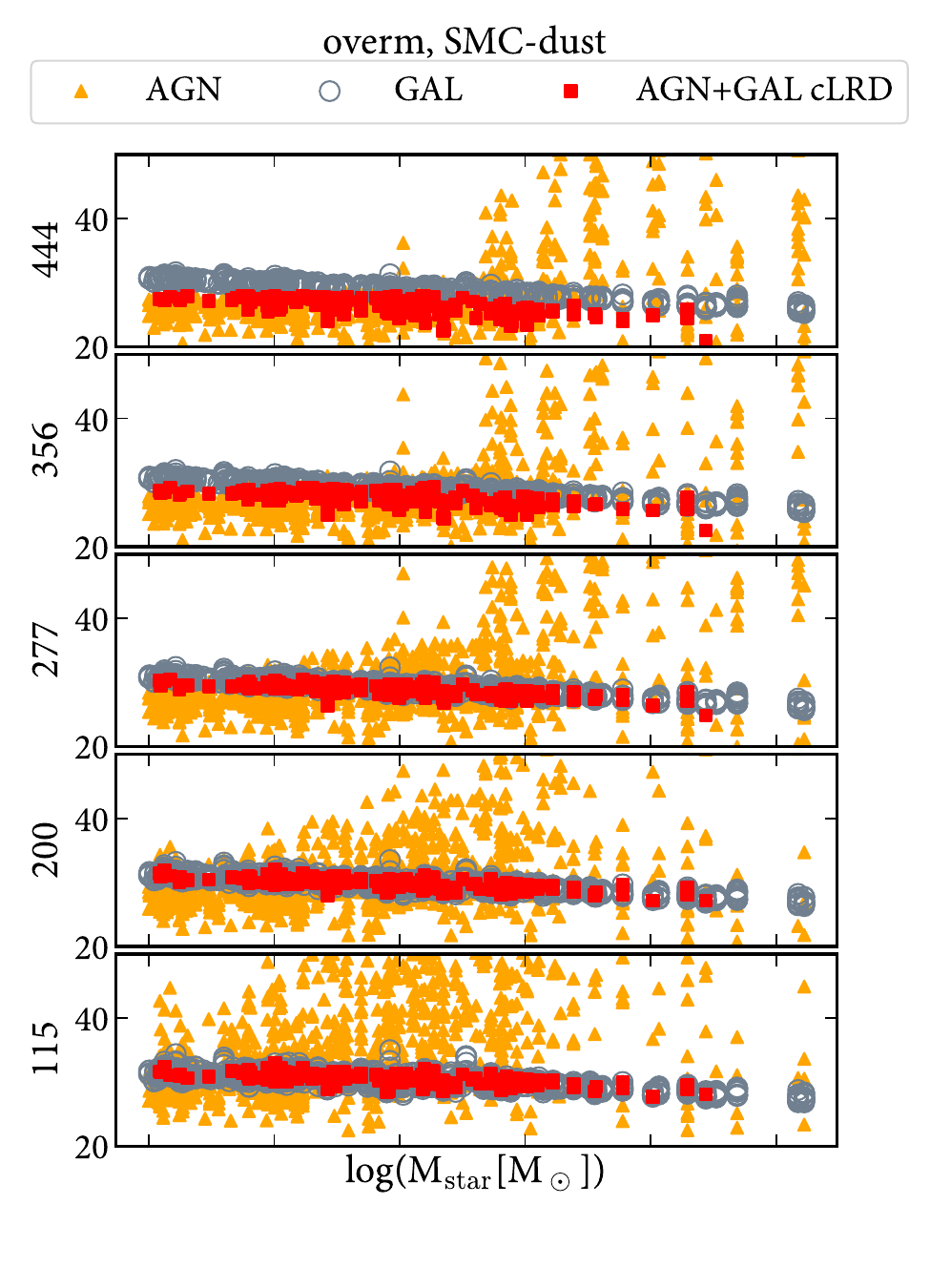}
\caption{Magnitudes for the `sEdd' and `overm' models and SMC-like dust.  The orange triangles and gray circles show separately the flux of AGN and galaxies, when selecting systems with $L_{\rm bol}>10^{44} \,{\rm erg \, s}^{-1}$ and m444W$<$28 (total flux, summing both AGN and galaxy contribution). The red dots have both of the above, plus the UNCOVER color selection, and show the combined galaxy+AGN flux. We have limited the y-axis to 50 to highlight the relevant parameter space. At 115 and 200 the galaxy is almost always brighter than the AGN, whose contribution increases as the wavelength increases. For the `overm' case some AGN are brighter than the galaxy also at short wavelenghts. }
              \label{fig:mag_z6_80}
    \end{figure*}

\section{Spectral energy distributions and absorption}
\label{sec:SED_abs}

At these redshifts, for our unattenuated models the galaxy dominates the emission at short JWST wavelengths, and the AGN at long JWST wavelengths \citep[see][for models without dust]{2017ApJ...849..155V,2023MNRAS.521..241V}, as long as the AGN is sufficiently bright with respect to the host galaxy. This can be intuitively understood as follows \citep[see Fig. 2 in][]{2017ApJ...849..155V}. In absolute terms, galaxies are star-forming and dominated by young stellar populations (Fig.~\ref{fig:galprops}), with a blue intrinsic spectrum that falls off more steeply than the AGN intrinsic SED at longer wavelengths. Furthermore,  AGN SEDs peak at lower frequencies as the MBH mass decreases, and the AGN emission dominates only when the MBH represents a sizeable fraction of the galaxy mass. The dominance of galaxy emission over the AGN at short JWST wavelengths is only accentuated when including attenuation.This is shown in Fig.~\ref{fig:SEDs} for some idealized cases. There is however a variety of behaviors, and we will show more realistic examples in Section~\ref{sec:los}. In the left panels of Fig.~\ref{fig:SEDs} we vary smoothly  \mbh{} and \fedd{} at fixed gas and dust column density seen by the AGN, and show random line of sight through the galaxy for the stellar population. In the right panels we fix \mbh{} and \fedd, vary smoothly the AGN column density, and show three different lines of sight for the stellar population. We can see that in this model the wavelength where the SEDs transition from the galaxy to AGN dominated changes as a function of dust model, column density and MBH properties.  Those that are selected as cLRDs have the transition in the ``right place'' to obtain a v-shaped SED. 

Fig.~\ref{fig:NH_size} shows the dependence of the mean hydrogen column density from the ISM only on galaxy mass and size. In general low-mass galaxies are diffuse because of efficient SN feedback puffing up the gas component (hence stars follow). While at larger masses ($\mstar \gtrsim 10^9\,\rm M_\odot$), SNe become inefficient, gas can pile up, and leads to wet compaction \citep{2014MNRAS.438.1870D}.   We first discuss galaxy sizes: 75\% of galaxies with $\mstar>10^9 \, \msun$ have half mass radius smaller than 300~pc (100\% for $\mstar>3\times 10^9 \, \msun$). The most compact galaxy in our sample has $R_{{\rm 50,mass}}=55$~pc for a mass of $\mstar=1.84 \times 10^9 \, \msun$. The `max' column density is clearly highly correlated to galaxy mass and compactness, while, obviously, removing a sphere with a radius of 80~pc makes the most compact galaxies less optically thick. This is particularly noticeable for galaxies with $R_{\rm 50,mass}<100$~pc. In Fig.~\ref{fig:NH_size} one can also appreciate the importance of the inhomogeneites in gas distribution: for a subset of galaxies we show the 12 lines of sight we use, and we see that for a given galaxy there can be a spread of 1-2 orders of magnitude. The ISM column density seen by the AGN increases with stellar mass as galaxies become more compact. A contribution of $10^{22}-10^{23}$~cm$^{-2}$ from the ISM to the total column density is in good agreement with the results obtained by \citet{2023ApJ...951L..41S} on 3 AGN at $z\sim 0.75-2$.  

In the bottom panels of Fig.~\ref{fig:NHdistr} we show `outmax80' AGN gas column densities for  models that produce more than 10\% of cLRDs (see Section~\ref{sec:colors}). The bifurcation at $M_{\rm star}>10^{10}\,\rm M_\odot$ is related to the ability of radiation pressure to lift gas: the high branch is where the column density is too large for radiation pressure on dust to lift dusty gas, the low branch is the opposite case. This is noticeable only at the high mass end because in most low mass galaxies the intrinsic column density is below the limit to counteract radiation pressure.  In general, most of our simulated AGN have gas column density above $10^{22} {\rm cm \, s^{-2}}$, therefore unobscured, UV-bright AGN are a minority, associated with favorable lines of sight. Simulations  typically find low average column densities only after the energy injection from the AGN has cleared out its surroundings \citep{2019MNRAS.487..819T,2020MNRAS.495.2135N,2022MNRAS.510.5760L}. However, since in Obelisk the MBHs have not grown much, they have injected little energy. 

The galaxy SED is modulated by the star formation history of the galaxy and by the gas and dust distribution. 
Attenuation for starlight is not the same at all wavelengths, but it depends on where stars are located: for instance, young stars dominating the emission at short wavelengths are typically embedded in dense gas clouds, and therefore they are surrounded by higher column densities. This can be appreciated in the top panels of Fig.~\ref{fig:NHdistr}, where we have rescaled the optical depth by the cross section and dust-to-gas ratio to obtain an effective column density:
\begin{equation}
N_{\rm H}=\frac{ X_{\rm H} \, \tau_\lambda}{\kappa_\lambda m_{\rm p} {\rm DTG}},   
\label{eq:effNH}
\end{equation}
where $\tau_\lambda$ is obtained from the intrinsic and attenuated fluxes at wavelength $\lambda$, $\kappa_\lambda$ is the absorption coefficient, $m_{\rm p}$ is the proton mass, $X_{\rm H}=0.76$ is the universal hydrogen mass fraction and DTG is the dust-to-gas mass ratio.  This effective column density can be compared to the AGN's, shown in the bottom panels. 
 The AGN is generally much more absorbed than starlight, since by definition it is located at the core of the galaxy, and this is the case for cLRDs as well, as shown by the red points in all panels. This also explains why the dust model, e.g., MW vs SMC has a stronger impact on the AGN compared to the galaxy. The ratio of attenuated luminosities for two dust models at fixed intrinsic luminosity is $L_{\rm att,dust1}/L_{\rm att,dust2}=\exp\left( N_{\rm H} {\rm DTG}(\kappa_{\rm dust2}-\kappa_{\rm dust1}) m_{\rm p}/X_{\rm H}\right)$. Since $N_{\rm H}$ is larger for AGN than for starlight, the difference between two dust models is stronger for AGN than for the stellar population.

The X-ray absorbed fraction, defined as the fraction with $N_{\rm H}>10^{23}\, {\rm cm}^{-2}$,  is shown in Fig.~\ref{fig:absfrac}. Except for the `max' model, which overpredicts the absorbed fraction, all other models underpredict the absorbed fraction from X-ray observations, unless we include the hydrogen column density within the sublimation radius.
We have not attempted to modify the X-ray emission to account for the lack of X-ray emission from some/many high-z AGN and red quasars \citep{Ma2024,2024ApJ...974L..26Y,2024ApJ...969L..18A, 2024arXiv240500504M}. Given that we find hydrogen column densities below $10^{25} \,{\rm cm}^{-2}$  the X-ray luminosity of the AGN at 2-10~keV observer's frame,  corresponding  to $\sim 10-40$~keV rest frame, would be dimmed by at most by 10\% for $L_{\rm 2-10keV}>10^{43} \,{\rm erg\, s^{-1}}$; this is small compared to the discrepancy seen in observations that is more than 2 orders of magnitudes for most JWST AGN. In the $L_{\rm 2-10keV}>12024Natur.636..594J0^{43} \,{\rm erg\, s^{-1}}$ luminosity range the typical total hydrogen column density is between $10^{22.5}$ and $10^{23.5}$~cm$^{-2}$. The luminosities of the simulated AGN reach $10^{45} \,{\rm erg\, s^{-1}}$, therefore the AGN would be detectable in deep X-ray observations.

To explain the lack of X-rays from high-z AGN one option is that X-ray and optical observations select intrinsically different populations. In this case, the X-ray bright population would be missed in optical searches because the AGN are too much reddened. However, the larger number density of broad-line AGN compared to that of X-ray selected AGN implies that the large majority of AGN is intrinsically X-ray weak \citep{Ma2024}, or that absorption is dominated by dust-free gas as suggested by \citet{2024arXiv240500504M}.  In our model, however, we do not reach column densities within the sublimation radius that would dim the X-rays at the rest frame energies: for  $\sim 10-40$~keV rest frame a column density $>10^{25} \, {\rm cm^{-2}}$ would be needed. Another possibility is that some AGN \citep[but not all, see e.g.,][]{2024NatAs.tmp..262S,2024MNRAS.528.1542M} in the super-Eddington regime are intrinsically X-ray weak, because of suppression of the production of UV photons that can be upscattered due to radiation trapping, or disc winds shielding or disrupting the corona or failed thermalization \citep{2005ApJ...630L...9P,2017MNRAS.464.1102B,2018ApJ...859L..20D,2019ApJ...880...67J,2024ApJ...976...96P}. This ties in particularly well with the `sEdd' case, and is in good agreement with the suggestion that many if not most of the high-z AGN detected by JWST are in the super-Eddington regime \citep{2023MNRAS.526.3250S,2024A&A...689A.128L,2024Natur.636..594J}.

\section{Photometric selection of Little Red Dots}
\label{sec:colors}

Our model does not include emission lines, therefore we start by comparing it with photometrically selected AGN. Specifically, we apply the color selection adopted in the UNCOVER\footnote{We show in the Appendix results obtained with the alternative color selection of cLRDs used in \citet{2024ApJ...968....4P}. } survey \citep{2025ApJ...978...92L,2024ApJ...964...39G}, plus a threshold in bolometric luminosity at $10^{44} \,{\rm erg\, s^{-1}}$ to mimic the H$\alpha$ flux/EW requirement. We do not impose a compactness criterion. For simplicity, we approximate the flux limit as  m444W$<$28, which is applied to the combined AGN+galaxy flux unless noted otherwise. The color selection requires the following criteria: $-0.5<{\rm m115W-m200W}<1.0$, ${\rm m277W-m444W}>1.0$, ${\rm m277W-m356W}>0.7$. We define as full AGN population systems with combined AGN+galaxy m444W$<$28 and bolometric luminosity $>10^{44} \,{\rm erg\, s^{-1}}$.  

We first discuss the galaxy properties, without including AGN, shown in Fig.~\ref{fig:colors_dust_nodust} (top and middle panels). The full galaxy population is shown with black points, galaxies brighter than m444W=28 in gray. None of the "galaxy-only" models get close to the cLRD classification, therefore if the cLRDs identified with this color selection are actually "not AGN", we still cannot explain their SED with our models. The same is true for models with no dust, with or without AGN (bottom panels).  Certainly, our model lacks emission lines and therefore galaxies that can be scattered into the selection thanks to the presence of strong emission lines. This is however not the most likely reason. \Obelisk{} galaxies are for the most part on the galaxy main sequence (Fig.~\ref{fig:galprops}, top) and do not have extreme ages, either very young or very old. The simulation therefore does not include very evolved stellar populations as in \citet{2024NatAs.tmp..284D}, or strong starbursts \citep{2024ApJ...968....4P,2024arXiv240610341A}. If the number density of non-AGN Little Red Dots is about $10^{-4} \, {\rm Mpc}^{-3}$ \citep{2024arXiv240610341A}, we should find ${\cal O}(1)$ of these galaxies in the \Obelisk{} volume, therefore it's not greatly surprising that we we do not find any. Finally, we have also assumed standard dust attenuation laws.

Moving on to the properties of AGN selected as cLRDs, in order to skew sufficiently the colors towards red, the AGN must not contribute much (or at all) at F115W and F200W, and should instead contribute, or even dominate, at F356W and especially at F444W. This can be achieved only if the AGN is intrinsically sufficiently luminous with respect to the host galaxy: it requires MBHs to be overmassive and/or have high Eddington ratios as noted in Fig.~\ref{fig:colors_z6_SMC}. As shown in Fig.~\ref{fig:colors_dust_nodust}, using the UNCOVER selection, no AGN in any of our models, including `overm' and `sEdd' enters the selection if dust is not included. When the AGN is too luminous with respect to the host galaxy it falls out of the UNCOVER selection, though, because it becomes too red in m115W-m200W, especially in the case with SMC-like dust. The presence of emission lines, which are not included in our model, can temper these results, as they can boost the brightness in some JWST bands more than in others, for instance H$\alpha$ at $z\sim 5$ can boost flux in F356W and/or F444W, and help enter the selection.

\begin{figure}
   \centering
   \includegraphics[width=\columnwidth]{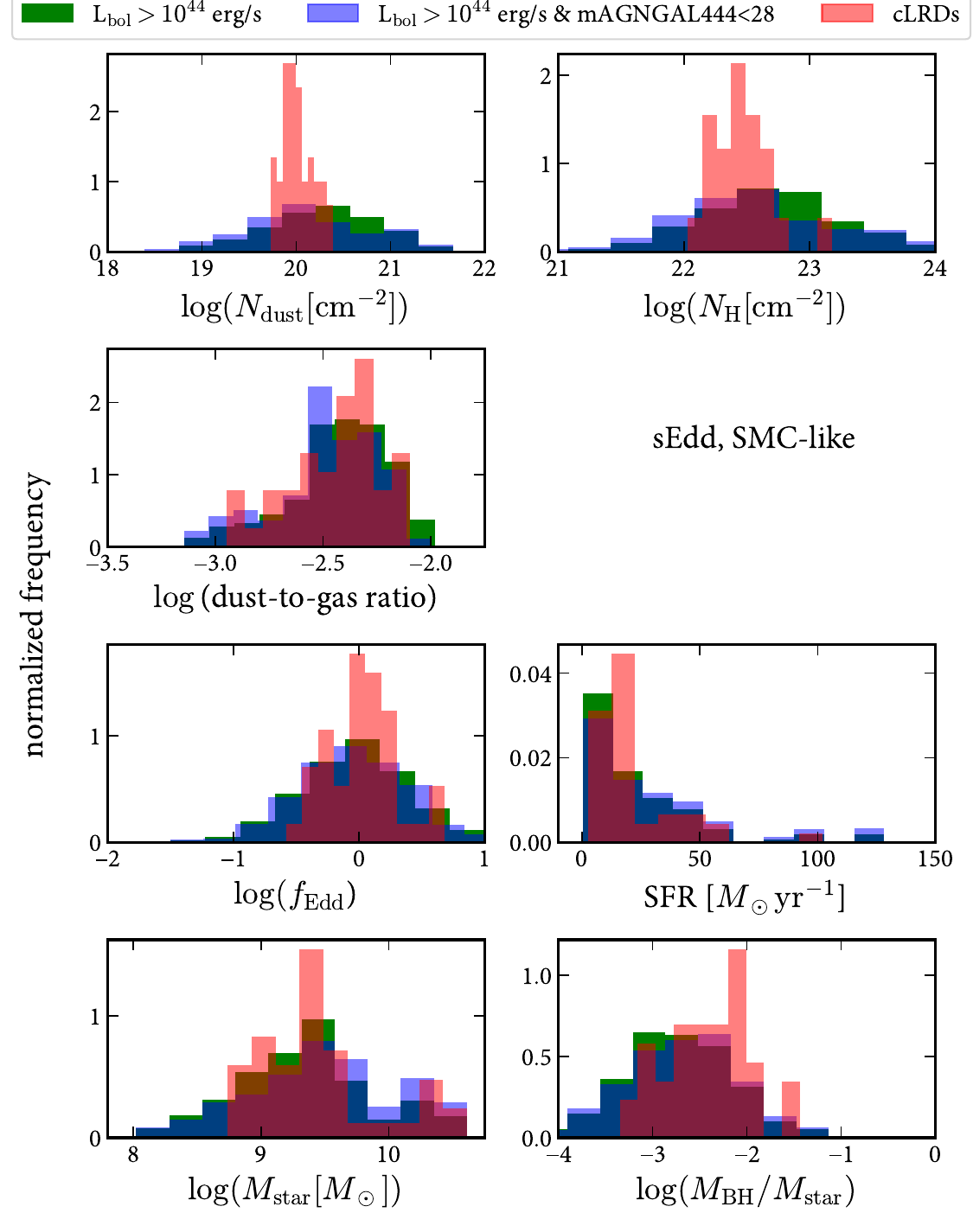}
    \caption{Properties of color-selected cLRDs (red) with respect to all AGN with $L_{\rm bol}>10^{44}\, {\rm erg \, s}^{-1}$ (green) and with further m444W$<28$ (blue) for SMC-like dust. Left: sEdd; right: overm. A clear signature is the specific range in dust column density, $N_{\rm dust}$, with respect to the global population. Further conditions are the elevated \fedd{} and ratio between \mbh{} and \mstar.}
              \label{fig:props_SMC_sEdd}
    \end{figure}

\begin{figure}
   \centering
   \includegraphics[width=\columnwidth]{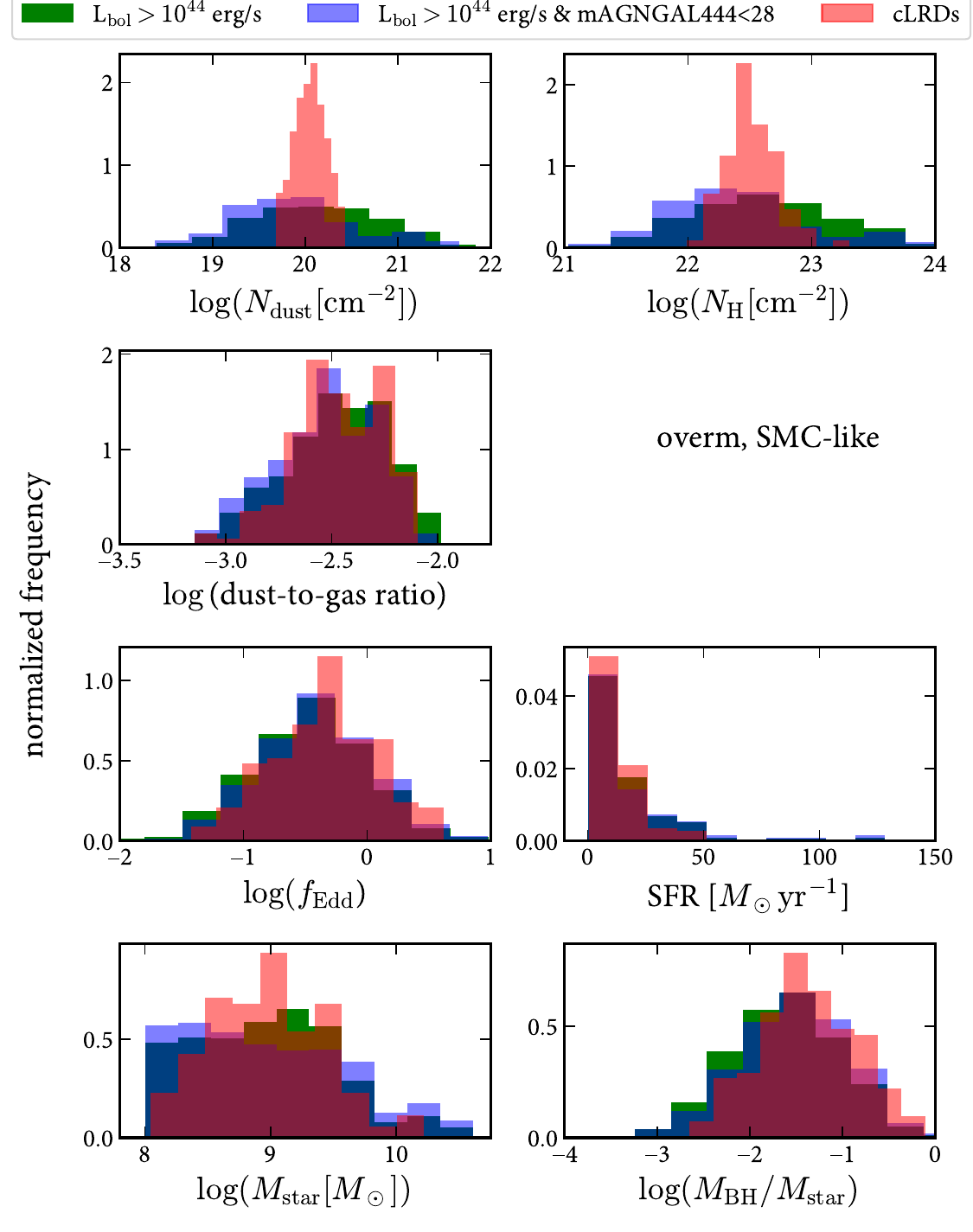}
    \caption{Properties of color-selected cLRDs (red) with respect to all AGN with $L_{\rm bol}>10^{44}\, {\rm erg \, s}^{-1}$ (green) and with further m444W$<28$ (blue) for SMC-like dust. Left: sEdd; right: overm. A clear signature is the specific range in dust column density, $N_{\rm dust}$, with respect to the global population. Further conditions are the elevated \fedd{} and ratio between \mbh{} and \mstar.}
              \label{fig:props_SMC_overm}
    \end{figure}

Keeping in mind the limitations of the simulation and of our model we have just discussed, the fact that the UNCOVER color selection, when applied to our model, is not picking galaxies, means that the UNCOVER color selection is well-suited to identify AGN, although it does not select the full AGN population: it recovers between 5\% and 26\% of all AGN with bolometric luminosity $L_{\rm bol}>10^{44} \,{\rm erg\,s^{-1}}$, depending on the assumed relation of MBH to stellar mass, AGN SED and dust properties, as shown in Fig.~\ref{fig:colors_z6_SMC}.  Based on \citet{2023ApJ...959...39H}, the red population accounts for $\sim 20\%$ of the broad-line selected AGN, but note that differing color selections were involved in spectroscopic targeting for different surveys. We will discuss this further in Section~\ref{sec:props}, but qualitatively, the selection misses AGN powered by MBHs with \mbh-\mstar{} ratio $<3 \times 10^{-2}$, low \fedd, as well as AGN with very high or very low column density. For the SMC-like dust case, a way to increase the completeness of the AGN selection is to extend the selection further red in ${\rm m115W-m200W}$ to $-0.5<{\rm m115W-m200W}<3.0$, leaving the rest of the selection criteria unchanged. Depending on the specific SED and dust model, the fraction of AGN in the selection increases by 50-100\%.  About half of the systems with 1<m115W-m200W<3 (plus the other UNCOVER cuts) are cases where the AGN already contributes at F200W, because of low column density (for the AGN) and/or low SFR (for the galaxy) and/or very high \mbh{} and \fedd{} that lead the AGN to dominate everywhere. The others are galaxies with steep spectrum and often low SFR.  MBHs with low \mbh-\mstar{} ratio and/or \fedd{}  typically fail to enter the selection because they are unable to contribute sufficiently to the emission, which is dominated by the galaxy starlight. This was already noted in \cite{2017ApJ...849..155V, 2023MNRAS.521..241V} for models that did not include dust. In the `overm' case, the intrinsic MBH population is already skewed towards large MBH-to-galaxy ratios, and therefore it is easier for MBHs to contribute significantly to, or dominate, the emerging flux, as can be appreciated in Fig.~\ref{fig:mag_z6_80} (right). In the `sEdd' case, instead, the redder peak of the SED increases the AGN flux in the red bands (redward of $\sim$ F227W), easing their inclusion in a red selection (Fig.~\ref{fig:mag_z6_80}, left).  

The MW-like dust model\footnote{All figures related to MW-like dust can be found at \href{https://doi.org/10.5281/zenodo.14628499}{10.5281/zenodo.14628499.}} generally leads to a smaller fraction of cLRDs than SMC-like dust, except in the `overm' case where the fraction instead increases. This is because fewer AGN have ${\rm m115W-m200W}>1$. Furthermore, with MW-like dust galaxies and especially AGN are fainter, therefore more are missed when selecting systems with m444W$<$28. The number of AGN at the denominator decreases significantly, thus increasing the fraction of cLRDs over the full AGN population. Finally, with MW-like dust, the AGN with ${\rm m115W-m200W}>1$ are a minority, $\sim$ 3\%.  

Finally, we have assumed the same dust properties for the galaxy component and for the AGN. There is the possibility to have internal regions of a galaxy with a dust type (e.g. SMC) and external regions with another type (e.g. MW). Hence, the dust type could be of one kind for AGN and another kind for the galaxy component.  If dust near AGN is SMC-like and dust in the rest of the galaxy is MW-like, cLRD selection is much more effective, and vice-versa. We may also expect variations in the extinction law for dust near AGN due to differential sublimation (smaller grains are more easily sublimated and have a larger sublimation radius). This generally translates into a flatter extinction curve \citep{2004ApJ...616..147G,2010A&A...523A..85G,2023Galax..11..102A}. We show an example in the Appendix, where we assume that dust for the stellar population is SMC-like and for the AGN is described by the empirical extinction curve from \citet{2004ApJ...616..147G}.

 \begin{figure}
   \centering
   \includegraphics[width=\columnwidth]{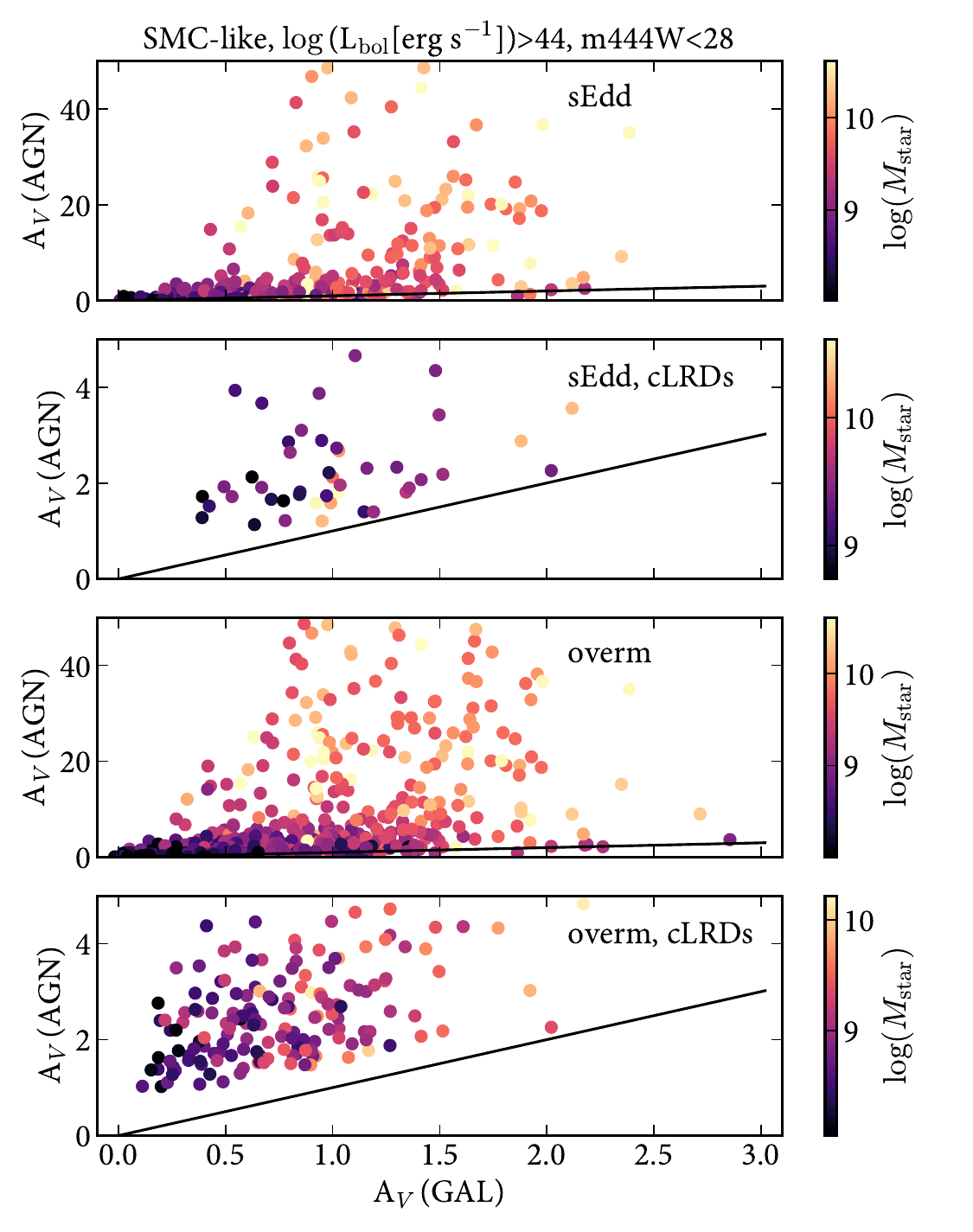}
    \caption{$A_V$ for AGN versus $A_V$  stellar population, for AGN with $L_{\rm bol}>10^{44}\, {\rm erg \, s}^{-1}$ and m444W$<28$ (first and third rows from top, where we have limited the y-axis to $A_V=50$) and for systems selected as cLRDs (second and fourth rows from top). We only show SMC-like dust for the sEdd and overm models.  $A_V$ for the stellar population of systems selected as cLRDs is moderately small with respect to the full extent of the distribution, while  $A_V$ for the AGN is significantly skewed towards low values with respect to the global population.}
              \label{fig:AV}
    \end{figure}

 \begin{figure*}
   \centering
   \includegraphics[width=0.48\textwidth]{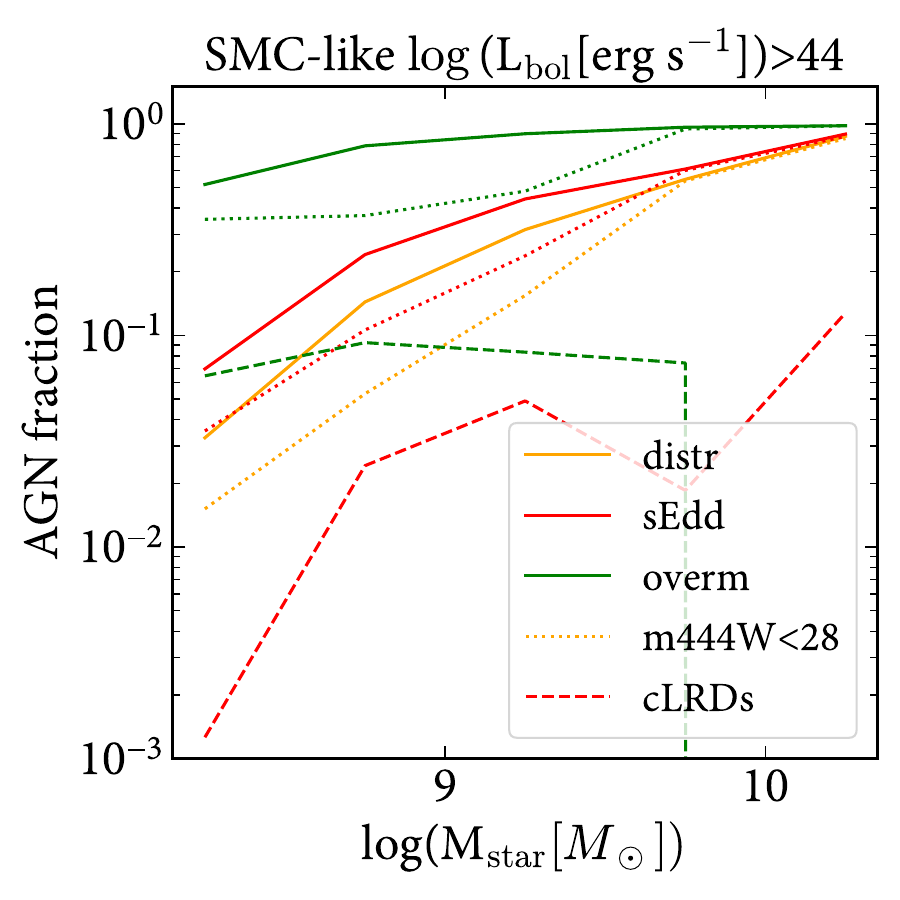}
   \includegraphics[width=0.48\textwidth]{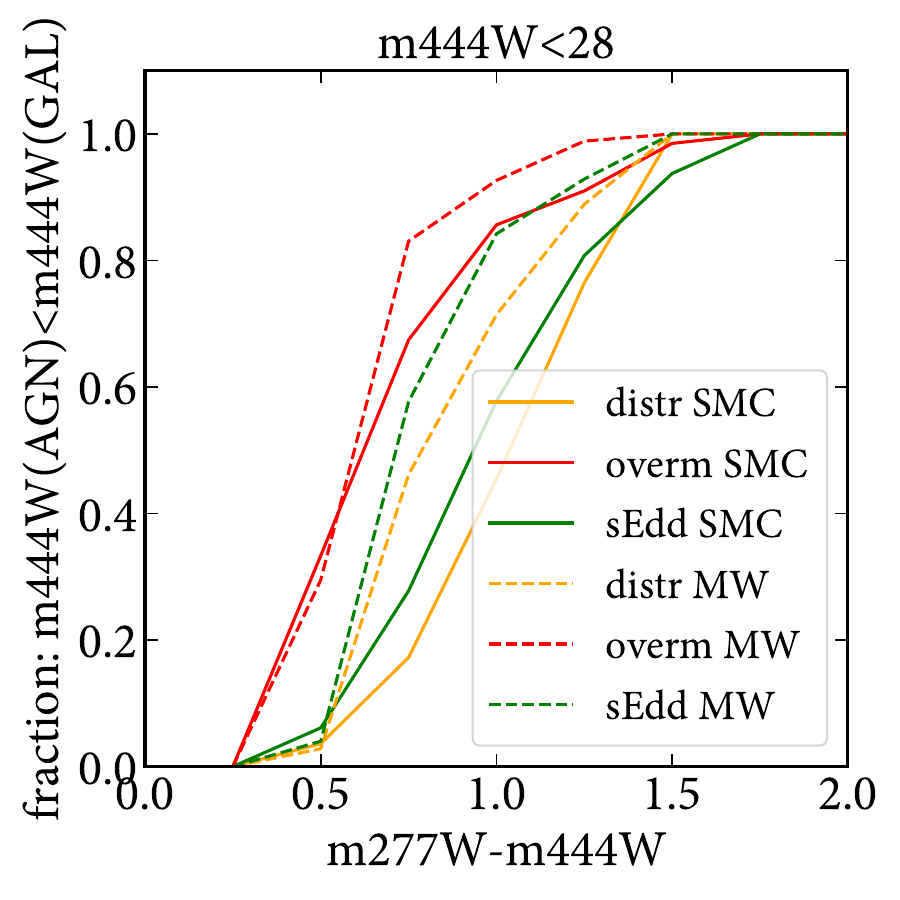}
    \caption{Left: AGN fraction, defined in terms of AGN bolometric luminosity as a function of galaxy mass. We recall that we have assumed an intrinsic active fraction of unity. Solid lines only require  $L_{\rm bol}>10^{44}\, {\rm erg \, s}^{-1}$, dotted lines add m444W$<28$ (galaxy and AGN combined) and dashed lines are for cLRDs (including the same bolometric and m444W cuts). Right: fraction of AGN-dominated sources at F444W as a function of m277W-m444W. Above m277W-m444W>1.5 all sources are AGN-dominated.}
              \label{fig:AGNfraction}
    \end{figure*}

  \begin{figure}
   \centering
   \includegraphics[width=\columnwidth]{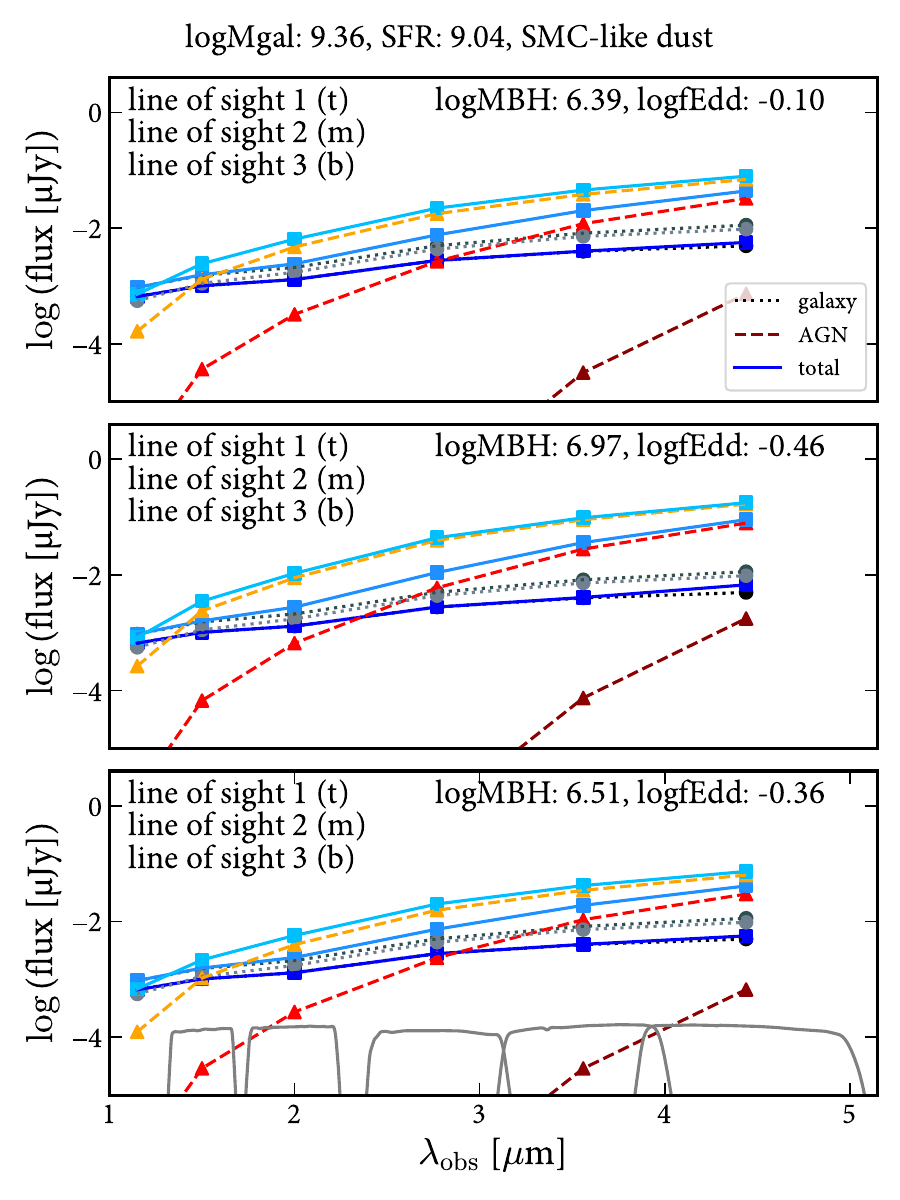}
     \caption{Examples of photometry at $z=6$ showing how different assumptions affect the emergent SEDs. For the same galaxy we show different \mbh, \fedd{} and lines of sight. The galaxy photometry is calculated along the same line of sight as that of the AGN. The SFR is in units of $M_\odot\,\mathrm{yr^{-1}}$ and masses in units of $M_\odot$ inside the logarithm. The NIRCAM response for the filters analyzed in this paper is shown in gray.}
              \label{fig:SEDs_LOS}
    \end{figure}

\section{Properties of AGN and color-selected Little Red Dots}
\label{sec:props} 

We compare the properties of cLRDs to the full MBH population and AGN with $L_{\rm bol}>10^{44}\, {\rm erg \, s}^{-1}$ and m444W$<28$ in Figs.~\ref{fig:props_SMC_sEdd} and~\ref{fig:props_SMC_overm}. The clearest distinctive feature of cLRDs is their intermediate dust column densities.  For high dust column densities the AGN is too buried and does not contribute sufficiently to the photometry. If the dust column density is too low, then AGN are not `red enough' as the dust-less case is not picked up by the selection, as shown in Fig.~\ref{fig:colors_dust_nodust}. 

The bimodal behavior in the dust and hydrogen column density when imposing m444W$<28$ (compare green and blue histograms) can be understood looking at Figs.~\ref{fig:NHdistr} and~\ref{fig:mag_z6_80}. In galaxies with $M_{\rm star}<10^{10}\,\rm M_\odot$, the intrinsic luminosities are smaller, therefore the magnitude cut selects galaxies (or lines of sight) with low column densities. In the most massive galaxies, the intrinsic luminosity is higher, therefore they can meet the magnitude cut even when column densities are higher: the high branch of the column density at $M_{\rm star}>10^{10}\,\rm M_\odot$. This combines with the relation between MBH and galaxy luminosities, favoring overmassive MBHs with relatively high \fedd. Even in the overm case, the median $\log(\mbh-\mstar)$ ratio is higher for cLRDs than for all AGN with $L_{\rm bol}>10^{44}\, {\rm erg \, s}^{-1}$: -1.4 for cLRDs vis-a-vis -1.6 for all AGN with $L_{\rm bol}>10^{44}\, {\rm erg \, s}^{-1}$. For the sEdd case we have -2.4 for cLRDs, and -2.7 for $L_{\rm bol}>10^{44}\, {\rm erg \, s}^{-1}$. The median \fedd=1.0 is the same for cLRDs and $L_{\rm bol}>10^{44}\, {\rm erg \, s}^{-1}$ AGN in the sEdd case, while it's -0.3 for cLRDs and -0.5 for $L_{\rm bol}>10^{44}\, {\rm erg \, s}^{-1}$ in the overm case. 

In Fig.~\ref{fig:NHdistr} we overlay in red the cLRDs to highlights how these points are interrelated. It is obviously easier for more massive galaxies to host more massive BHs, which at fixed \fedd{} will more easily meet the bolometric luminosity requirement. However, if low-mass galaxies host sufficiently massive MBHs to power a bright AGN, they are "easier" hosts for cLRDs, because low-mass galaxies have lower column densities compared to their high-mass counterparts. The AGN can then more easily contribute to the total emission and skew colors.  We also recall that in the `sEdd' case the redder intrinsic AGN SED helps to meet the color selection. 

 \begin{figure}
   \centering
   \includegraphics[width=\columnwidth]{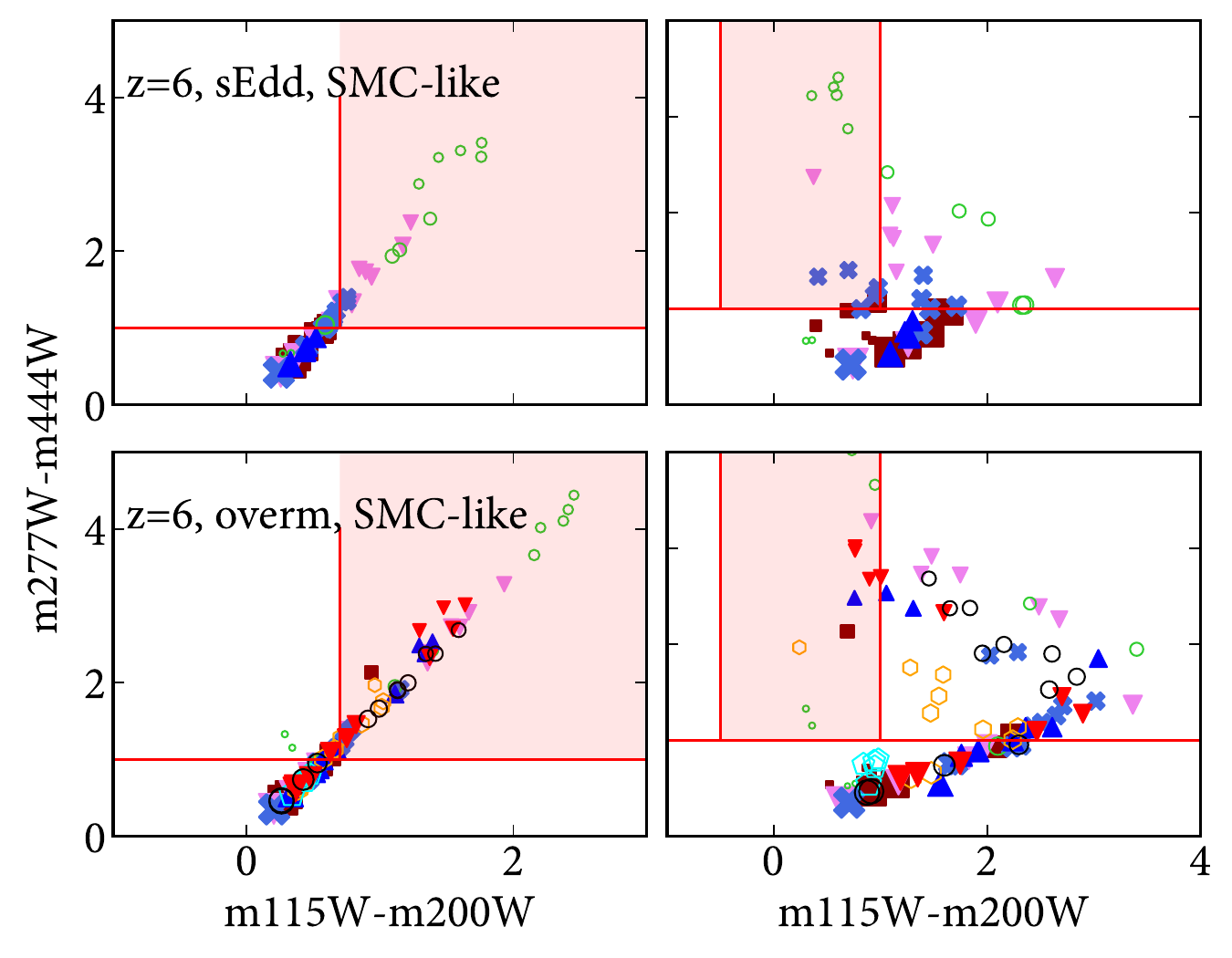}
     \caption{Colors for 9 galaxies where we explore how the appearance changes over 12 lines of sight, fixing \mbh{} and \fedd. We show the sEdd and overm cases, for SMC-like dust. Each galaxy is shown with a different combination of color and symbol, therefore multiple identical markers indicate different lines of sight for the same galaxy+AGN intrinsic properties. The size of the symbol scales with the AGN optical depth. We show only cases with m444W<28 (combined galaxy+AGN), but we do not impose any threshold on the AGN luminosity.}
              \label{fig:colors_LOS}
    \end{figure}

In terms of dust-to-gas ratio and SFR, color-selected AGN are evenly distributed within the global population.  \Obelisk{} galaxies are generally actively star-forming and have dust-to-gas ratios between $10^{-3}$ and $10^{-2}$, and cLRDs span the same range, with a preference for low SFR. Certainly \Obelisk{} galaxies have formed enough dust to explain the observed colors. 

A question that often arises is what is the attenuation of cLRDs in terms of $A_V$. This is shown in Fig.~\ref{fig:AV} for the `overm' and `sEdd' cases.  When considering only cLRDs, it is immediately obvious that they preferentially occupy a narrow range of $A_V=$1--5. This is because some attenuation is needed to meet the color selection, but enough AGN light has to escape in order to affect the colors. The `overm' case, which has the brightest AGN at fixed galaxy mass, allows for somewhat larger $A_V{\rm(AGN)}$, followed by the `sEdd' case. The median AGN $A_V$ is 2.4, while for the `sEdd' case it's 2.1. These values are somewhat higher than the median 1.6 found by \citet{2024ApJ...968...38K} and somewhat lower than the typical value of 3 found by \citet{2024arXiv240610341A}.

Importantly, cLRDs are not the most attenuated sources in our analysis. Attenuation of the AGN component when we do not apply a magnitude cut can reach extreme values of 50 mag. These are cases that obviously preclude the AGN from showing any photometric signature. This is in contrast with the attenuation for the galaxy component, for which $A_V$ is always limited to 3 mag. The median $A_V$ for the stellar population is between 0.8 (sEdd) and 0.9 (overm). This is a consequence of the different (dust) column densities for the MBH and the stellar population. Fig.~\ref{fig:NHdistr} highlights how NH for the stellar population, modulated by the location of young and old stars, is generally an order of magnitude smaller than for the centrally located MBH. The optical depth is proportional to the column density, and $A_V$ is proportional to the optical depth. An order of magnitude larger column density therefore translates into an order of magnitude larger $A_V$.   

In Fig.~\ref{fig:AGNfraction} (left) we turn to the global AGN population showing the AGN fraction as a function of stellar mass. The intrinsic active fraction, meaning the probability that a galaxy hosts an active MBH is unity in the model, but here we adopt an observer's point of view, and consider the fraction of AGN bright enough to be detected, assuming that $L_{\rm bol}>10^{44}\, \rm erg\, s^{-1}$ is necessary to power sufficiently broad and bright emission lines. In all cases the fraction of AGN with $L_{\rm bol}>10^{44}\, \rm erg\, s^{-1}$ is an increasing function of \mstar. This is because we have assumed a mass-independent \fedd{} distribution, while MBH mass increases with galaxy mass. The `overm' case has significantly higher MBH masses, therefore the AGN fraction remains high also in low-mass galaxies.  Adding the condition m444W$<28$, set on the combined AGN+galaxy, decreases the AGN fraction in low-mass galaxies. This is because in  galaxies with mass $\gtrsim 3\times 10^{10} \, \msun $ the stellar population by itself can have m444W$<28$. This can be appreciated in Fig.~\ref{fig:mag_z6_80}: at the high-mass end cases with m444W of the AGN as faint at 40-50 can still meet the condition m444W$<28$ on the total flux.  Finally, the fraction of cLRDs is more sensitive to the MBH population properties (sEdd vs overm; we do not show the distr case because too noisy) in galaxies with mass $\lesssim 10^{9} \, \msun $. In the overm case it is easier for the brighter AGN to dominate on the galaxy colors. 

In the right panel of Fig.~\ref{fig:AGNfraction} we touch on the observed compactness of sources. If we assumed that compactness meant AGN-dominated at F444W, then we find that sources with m277W-m444W>1.5, have the AGN dominating photometry at F444W, i.e., m444W(AGN)<m444W(GAL). This agrees with the very high broad-line fractions and very high compact fractions found in observations for the same color cut. In other words, the redder the 277-444 color, the more AGN dominated is the light at F444W. We stress, however, that since we have no galaxy that without an AGN component has m277W-m444W>1.5, and galaxies are found with such colors \citep{2024ApJ...968....4P}, we cannot conclusively say that all sources with m277W-m444W>1.5 are only AGN \citep[see the discussion in][]{2024arXiv240610341A}. 
 
\section{Dependence on the line of sight}
\label{sec:los}

The importance of the dust column density in determining whether an AGN is a cLRD leads us to explore the effect of observing a galaxy at fixed MBH and macro galaxy properties, through different lines of sight. In this section instead of drawing different \mbh{} and \fedd{} for each line of sight, we use the same \mbh{} and \fedd{} for all 12 lines of sight of a given galaxy. 

In Fig.~\ref{fig:SEDs_LOS} we show examples of the photometry for a galaxy, where for different \mbh{} and \fedd{} we show three different lines of sight. Fixing \mbh{} and \fedd{} we can have, for the same source, an AGN-dominated case, a galaxy-dominated case and an intermediate case. 

The AGN dust column densities differ by 0.3 and 0.84~dex between the three lines of sight, while the effective dust column density for the stellar populations (Eq.~\ref{eq:effNH}) differ at most by 0.1~dex. The line-of-sight difference is therefore much more pronounced for the AGN. This is because it is more sensitive to the position of few cells.  

We have repeated the analysis shown in Fig.~\ref{fig:colors_z6_SMC} and find the following cLRD fractions $N({\rm cLRDs})/N({\rm AGN})$: 9\% (distr), 13\% (sEdd) and 18\% (overm) for SMC-like dust and 4\% (distr), 13\% (sEdd) and 22\% (overm) for MW-like dust, compatible with those obtained varying \mbh{} and \fedd{} for each line of sight. As discussed in the previous section the dust column density is as important as, if not more than, the intrinsic galaxy and MBH properties in determining whether an AGN can be classified as a cLRD.  

It is instructive to see how at fixed galaxy and MBH properties different lines of sight move sources in the UNCOVER color selection. This is shown in Fig.~\ref{fig:colors_LOS} for 9 galaxies: three with mass $<10^{8.5} \msun$, four with mass $10^{8.5} -10^{9.5} \msun $ and two with mass $>10^{10} \msun$. The results can be understood in light of Figs.~\ref{fig:SEDs}, \ref{fig:SEDs_LOS} and~\ref{fig:colors_dust_nodust}. When the AGN optical depth is high (large symbols) the AGN is too obscured and the source has colors typical of the stellar population, failing therefore to enter the UNCOVER selection being too blue in m277W-m444W and in m277W-m356W. When the AGN optical depth is low (small symbols) the AGN is unabsorbed, resulting also in colors that are too blue in m277W-m444W and in m277W-m356W. Only for intermediate optical depth (medium-size symbols) the AGN contribution blueward of F356W is suppressed resulting in redder colors and meeting the selection criteria.

\section{Conclusions}

We have developed a comprehensive model of AGN photometry, including spatially-resolved attenuation, in galaxies in the \Obelisk{} simulation at $z=6$. We first validate our model against observations, namely the AGN bolometric luminosity function and the X-ray obscured fraction. We then select AGN and cLRDs and investigate their properties. Our main results are as follows. 

\begin{itemize}
    \item Using realistic galaxy and AGN SEDs, with SMC- or MW-like dust we find that, as long as the AGN is sufficiently luminous with respect to the host galaxy, the AGN dominates in red NIRCAM filters and the galaxy in blue  NIRCAM filters.
    \item AGN span a variety of gas and dust column densities, which depend both on the physical properties of the host galaxy and on the line of sight as fixed properties. 
    \item The ISM contributes significantly to the total column density, and the contribution increases with galaxy mass and decreases with galaxy size.
    \item The simulation does not include any galaxy whose stellar population puts it in the Little Red Dot UNCOVER color selection.
    \item Colors effectively select AGN with intermediate dust column density/attenuation. When attenuation is too low or too high AGN do not meet the m277W-m444W criterion, being too blue, and are not selected as cLRDs.
    \item cLRDs have high Eddington ratios and/or \mbh-\mstar{} ratios. The host galaxies are within the general population, with only slightly lower SFR and not extreme dust-to-gas ratios.   
    \item Modulating the SED at super-Eddington rates to mimic radiation trapping, thus making the SED redder, eases the selection of AGN as cLRDs. Alternatively, this means that cLRD selection favors super-Eddington accretors. 
    \item cLRD selection is most sensitive to the MBH mass and accretion properties in galaxies with $\mstar<10^9 \msun$. 
    \item Sources with m277W-m444W>1.5 are AGN-dominated, which can lead to such sources appearing compact, in agreement with observations.
    \item Given that only intermediate attenuation gives rise to cLRDs, there is a significant dependence on the line-of-sight properties, meaning that theoretical models require spatially-resolved galaxy properties. 
    
\end{itemize}

Our analysis allows us to situate AGN selected as cLRDs within the AGN population, and understand their properties. Their spectral properties can be explained by AGN dominating in red filters and the host galaxy in blue filters because of intermediate attenuation and AGN that are intrinsically of luminosity comparable to that of the host galaxy. 

Super-Eddington accretion, via trapping of UV radiation close to the inner edge of the accretion disc, can explain better the optical spectral shape of cLRDs (more power in red vs blue) and there are arguments why it could be (but not always!) X-ray weak as many of the high-z AGN seem to be. In principle most of the high-z AGN could be super-Eddington. For cLRDs, super-Eddington helps pushing them towards the red selection, and therefore super-Eddington sources would naturally be more numerous with this selection. It is however unclear why purely spectroscopically selected AGN may also be predominantly super-Eddington. A possible explanation may be that super-Eddington accretion increases mass and luminosity of MBHs otherwise hosted in low-mass galaxies. Since low-mass galaxies are more numerous it would make it easier to preferentially select such sources \citep{2022MNRAS.511..616T}.

We have not found in the \Obelisk{} simulation any galaxy that alone, without a AGN component, enters the cLRD selections. This is partly caused by the lack of emission lines in our model, which can boost photometry in the filter where they fall. 

Based on these findings, future models need to account for spatially-resolved line-of-sight attenuation properties, and model AGN SEDs at with the same attention for physical properties as currently done for galaxies. It is also important to understand if/how galaxies can enter the cLRD selection, and how common such galaxies are in theoretical models.  

\section*{Data availability}
All figures related to MW-like dust can be found at \href{https://doi.org/10.5281/zenodo.14628499}{10.5281/zenodo.14628499.}

\begin{acknowledgements}
MV thanks Christina Eilers for clarifying the definition and usage of `Little Red Dots'. MT and PD acknowledges support from the NWO grant 0.16.VIDI.189.162  (``ODIN'') and PD acknowledges support from European Commission's and University of Groningen's CO-FUND Rosalind Franklin program. RS acknowledges support from  the PRIN 2022 MUR project 2022CB3PJ3 - First Light And Galaxy aSsembly (FLAGS) funded by the European Union - Next Generation EU.  This research was supported in part by grant NSF PHY-2309135 to the Kavli Institute for Theoretical Physics (KITP). This work has made extensive use of the Infinity Cluster hosted by Institut d'Astrophysique de Paris; we warmly thank Stéphane Rouberol for running smoothly this cluster for us. We acknowledge PRACE for awarding us access to Joliot Curie at GENCI@CEA, France, which was used to run most of the simulations presented in this work. Additionally, this work was granted access to the HPC resources of CINES under allocations A0040406955 and A0040407637 made by GENCI.
\end{acknowledgements}

%
\bibliography{biblio} 

\begin{thebibliography}{85}
\expandafter\ifx\csname natexlab\endcsname\relax\def\natexlab#1{#1}\fi

\bibitem[{{Adamo} {et~al.}(2024){Adamo}, {Atek}, {Bagley}, {Ba{\~n}ados},
  {Barrow}, {Berg}, {Bezanson}, {Brada{\v{c}}}, {Brammer}, {Carnall},
  {Chisholm}, {Coe}, {Dayal}, {Eisenstein}, {Eldridge}, {Ferrara}, {Fujimoto},
  {de Graaff}, {Habouzit}, {Hutchison}, {Kartaltepe}, {Kassin}, {Kriek},
  {Labb{\'e}}, {Maiolino}, {Marques-Chaves}, {Maseda}, {Mason}, {Matthee},
  {McQuinn}, {Meynet}, {Naidu}, {Oesch}, {Pentericci},
  {P{\'e}rez-Gonz{\'a}lez}, {Rigby}, {Roberts-Borsani}, {Schaerer}, {Shapley},
  {Stark}, {Stiavelli}, {Strom}, {Vanzella}, {Wang}, {Wilkins}, {Williams},
  {Willott}, {Wylezalek}, \& {Nota}}]{2024arXiv240521054A}
{Adamo}, A., {Atek}, H., {Bagley}, M.~B., {et~al.} 2024, arXiv e-prints,
  arXiv:2405.21054

\bibitem[{{Akins} {et~al.}(2024){Akins}, {Casey}, {Lambrides}, {Allen},
  {Andika}, {Brinch}, {Champagne}, {Cooper}, {Ding}, {Drakos}, {Faisst},
  {Finkelstein}, {Franco}, {Fujimoto}, {Gentile}, {Gillman}, {Gozaliasl},
  {Harish}, {Hayward}, {Hirschmann}, {Ilbert}, {Kartaltepe}, {Kocevski},
  {Koekemoer}, {Kokorev}, {Liu}, {Long}, {McCracken}, {McKinney}, {Onoue},
  {Paquereau}, {Renzini}, {Rhodes}, {Robertson}, {Shuntov}, {Silverman},
  {Tanaka}, {Toft}, {Trakhtenbrot}, {Valentino}, \&
  {Zavala}}]{2024arXiv240610341A}
{Akins}, H.~B., {Casey}, C.~M., {Lambrides}, E., {et~al.} 2024, arXiv e-prints,
  arXiv:2406.10341

\bibitem[{{Ananna} {et~al.}(2024){Ananna}, {Bogd{\'a}n}, {Kov{\'a}cs},
  {Natarajan}, \& {Hickox}}]{2024ApJ...969L..18A}
{Ananna}, T.~T., {Bogd{\'a}n}, {\'A}., {Kov{\'a}cs}, O.~E., {Natarajan}, P., \&
  {Hickox}, R.~C. 2024, \apjl, 969, L18

\bibitem[{{Antonucci}(2023)}]{2023Galax..11..102A}
{Antonucci}, R. R.~J. 2023, Galaxies, 11, 102

\bibitem[{{Begelman} \& {Volonteri}(2017)}]{2017MNRAS.464.1102B}
{Begelman}, M.~C. \& {Volonteri}, M. 2017, \mnras, 464, 1102

\bibitem[{{Bezanson} {et~al.}(2024){Bezanson}, {Labbe}, {Whitaker}, {Leja},
  {Price}, {Franx}, {Brammer}, {Marchesini}, {Zitrin}, {Wang}, {Weaver},
  {Furtak}, {Atek}, {Coe}, {Cutler}, {Dayal}, {van Dokkum}, {Feldmann},
  {F{\"o}rster Schreiber}, {Fujimoto}, {Geha}, {Glazebrook}, {de Graaff},
  {Greene}, {Juneau}, {Kassin}, {Kriek}, {Khullar}, {Maseda}, {Mowla},
  {Muzzin}, {Nanayakkara}, {Nelson}, {Oesch}, {Pacifici}, {Pan}, {Papovich},
  {Setton}, {Shapley}, {Smit}, {Stefanon}, {Taylor}, \&
  {Williams}}]{2024ApJ...974...92B}
{Bezanson}, R., {Labbe}, I., {Whitaker}, K.~E., {et~al.} 2024, \apj, 974, 92

\bibitem[{{Dai} {et~al.}(2018){Dai}, {McKinney}, {Roth}, {Ramirez-Ruiz}, \&
  {Miller}}]{2018ApJ...859L..20D}
{Dai}, L., {McKinney}, J.~C., {Roth}, N., {Ramirez-Ruiz}, E., \& {Miller},
  M.~C. 2018, \apjl, 859, L20

\bibitem[{{de Graaff} {et~al.}(2024){de Graaff}, {Setton}, {Brammer}, {Cutler},
  {Suess}, {Labb{\'e}}, {Leja}, {Weibel}, {Maseda}, {Whitaker}, {Bezanson},
  {Boogaard}, {Cleri}, {De Lucia}, {Franx}, {Greene}, {Hirschmann}, {Matthee},
  {McConachie}, {Naidu}, {Oesch}, {Price}, {Rix}, {Valentino}, {Wang}, \&
  {Williams}}]{2024NatAs.tmp..284D}
{de Graaff}, A., {Setton}, D.~J., {Brammer}, G., {et~al.} 2024, Nature
  Astronomy [\eprint[arXiv]{2404.05683}]

\bibitem[{{Dekel} \& {Burkert}(2014)}]{2014MNRAS.438.1870D}
{Dekel}, A. \& {Burkert}, A. 2014, \mnras, 438, 1870

\bibitem[{{Done} {et~al.}(2012){Done}, {Davis}, {Jin}, {Blaes}, \&
  {Ward}}]{2012MNRAS.420.1848D}
{Done}, C., {Davis}, S.~W., {Jin}, C., {Blaes}, O., \& {Ward}, M. 2012, \mnras,
  420, 1848

\bibitem[{{Dong} {et~al.}(2012){Dong}, {Greene}, \& {Ho}}]{Dong2012}
{Dong}, R., {Greene}, J.~E., \& {Ho}, L.~C. 2012, \apj, 761, 73

\bibitem[{{Dong-P{\'a}ez} {et~al.}(2023{\natexlab{a}}){Dong-P{\'a}ez},
  {Volonteri}, {Beckmann}, {Dubois}, {Mangiagli}, {Trebitsch}, {Vergani}, \&
  {Webb}}]{2023A&A...676A...2D}
{Dong-P{\'a}ez}, C.~A., {Volonteri}, M., {Beckmann}, R.~S., {et~al.}
  2023{\natexlab{a}}, \aap, 676, A2

\bibitem[{{Dong-P{\'a}ez} {et~al.}(2023{\natexlab{b}}){Dong-P{\'a}ez},
  {Volonteri}, {Beckmann}, {Dubois}, {Trebitsch}, {Mangiagli}, {Vergani}, \&
  {Webb}}]{2023A&A...673A.120D}
{Dong-P{\'a}ez}, C.~A., {Volonteri}, M., {Beckmann}, R.~S., {et~al.}
  2023{\natexlab{b}}, \aap, 673, A120

\bibitem[{{Dubois} {et~al.}(2013){Dubois}, {Pichon}, {Devriendt}, {Silk},
  {Haehnelt}, {Kimm}, \& {Slyz}}]{Dubois2013}
{Dubois}, Y., {Pichon}, C., {Devriendt}, J., {et~al.} 2013, \mnras, 428, 2885

\bibitem[{{Dubois} {et~al.}(2014){Dubois}, {Pichon}, {Welker}, {Le Borgne},
  {Devriendt}, {Laigle}, {Codis}, {Pogosyan}, {Arnouts}, {Benabed}, {Bertin},
  {Blaizot}, {Bouchet}, {Cardoso}, {Colombi}, {de Lapparent}, {Desjacques},
  {Gavazzi}, {Kassin}, {Kimm}, {McCracken}, {Milliard}, {Peirani}, {Prunet},
  {Rouberol}, {Silk}, {Slyz}, {Sousbie}, {Teyssier}, {Tresse}, {Treyer},
  {Vibert}, \& {Volonteri}}]{Dubois2014c}
{Dubois}, Y., {Pichon}, C., {Welker}, C., {et~al.} 2014, \mnras, 444, 1453

\bibitem[{{Dubois} {et~al.}(2024){Dubois}, {Rodr{\'\i}guez Montero}, {Guerra},
  {Trebitsch}, {Han}, {Beckmann}, {Yi}, {Lewis}, \&
  {Jang}}]{2024A&A...687A.240D}
{Dubois}, Y., {Rodr{\'\i}guez Montero}, F., {Guerra}, C., {et~al.} 2024, \aap,
  687, A240

\bibitem[{{Eldridge} {et~al.}(2017){Eldridge}, {Stanway}, {Xiao}, {McClelland},
  {Taylor}, {Ng}, {Greis}, \& {Bray}}]{2017PASA...34...58E}
{Eldridge}, J.~J., {Stanway}, E.~R., {Xiao}, L., {et~al.} 2017, \pasa, 34, e058

\bibitem[{{Elvis} {et~al.}(2002){Elvis}, {Marengo}, \&
  {Karovska}}]{2002ApJ...567L.107E}
{Elvis}, M., {Marengo}, M., \& {Karovska}, M. 2002, \apjl, 567, L107

\bibitem[{{Fabian} {et~al.}(2008){Fabian}, {Vasudevan}, \&
  {Gandhi}}]{2008MNRAS.385L..43F}
{Fabian}, A.~C., {Vasudevan}, R.~V., \& {Gandhi}, P. 2008, \mnras, 385, L43

\bibitem[{{Fabian} {et~al.}(2009){Fabian}, {Vasudevan}, {Mushotzky}, {Winter},
  \& {Reynolds}}]{2009MNRAS.394L..89F}
{Fabian}, A.~C., {Vasudevan}, R.~V., {Mushotzky}, R.~F., {Winter}, L.~M., \&
  {Reynolds}, C.~S. 2009, \mnras, 394, L89

\bibitem[{{Fan} {et~al.}(2001){Fan}, {Strauss}, {Richards}, {Newman}, {Becker},
  {Schneider}, {Gunn}, {Davis}, {White}, {Lupton}, {Anderson}, {Annis},
  {Bahcall}, {Brunner}, {Csabai}, {Doi}, {Fukugita}, {Hennessy}, {Hindsley},
  {Ivezi{\'c}}, {Knapp}, {McKay}, {Munn}, {Pier}, {Szalay}, \&
  {York}}]{2001AJ....121...31F}
{Fan}, X., {Strauss}, M.~A., {Richards}, G.~T., {et~al.} 2001, \aj, 121, 31

\bibitem[{{Gallerani} {et~al.}(2010){Gallerani}, {Maiolino}, {Juarez}, {Nagao},
  {Marconi}, {Bianchi}, {Schneider}, {Mannucci}, {Oliva}, {Willott}, {Jiang},
  \& {Fan}}]{2010A&A...523A..85G}
{Gallerani}, S., {Maiolino}, R., {Juarez}, Y., {et~al.} 2010, \aap, 523, A85

\bibitem[{{Gaskell} {et~al.}(2004){Gaskell}, {Goosmann}, {Antonucci}, \&
  {Whysong}}]{2004ApJ...616..147G}
{Gaskell}, C.~M., {Goosmann}, R.~W., {Antonucci}, R. R.~J., \& {Whysong}, D.~H.
  2004, \apj, 616, 147

\bibitem[{{Greene} {et~al.}(2024){Greene}, {Labbe}, {Goulding}, {Furtak},
  {Chemerynska}, {Kokorev}, {Dayal}, {Volonteri}, {Williams}, {Wang}, {Setton},
  {Burgasser}, {Bezanson}, {Atek}, {Brammer}, {Cutler}, {Feldmann}, {Fujimoto},
  {Glazebrook}, {de Graaff}, {Khullar}, {Leja}, {Marchesini}, {Maseda},
  {Matthee}, {Miller}, {Naidu}, {Nanayakkara}, {Oesch}, {Pan}, {Papovich},
  {Price}, {van Dokkum}, {Weaver}, {Whitaker}, \&
  {Zitrin}}]{2024ApJ...964...39G}
{Greene}, J.~E., {Labbe}, I., {Goulding}, A.~D., {et~al.} 2024, \apj, 964, 39

\bibitem[{{Greene} {et~al.}(2020){Greene}, {Strader}, \&
  {Ho}}]{2020ARA&A..58..257G}
{Greene}, J.~E., {Strader}, J., \& {Ho}, L.~C. 2020, \araa, 58, 257

\bibitem[{{Harikane} {et~al.}(2023){Harikane}, {Zhang}, {Nakajima}, {Ouchi},
  {Isobe}, {Ono}, {Hatano}, {Xu}, \& {Umeda}}]{2023ApJ...959...39H}
{Harikane}, Y., {Zhang}, Y., {Nakajima}, K., {et~al.} 2023, \apj, 959, 39

\bibitem[{{Hayes} {et~al.}(2024){Hayes}, {Tan}, {Ellis}, {Young}, {Cammelli},
  {Singh}, {Runnholm}, {Saxena}, {Lunnan}, {Keller}, {Monaco}, {Laporte}, \&
  {Melinder}}]{2024ApJ...971L..16H}
{Hayes}, M.~J., {Tan}, J.~C., {Ellis}, R.~S., {et~al.} 2024, \apjl, 971, L16

\bibitem[{{Jarvis} \& {MacAlpine}(1998)}]{1998AJ....116.2624J}
{Jarvis}, R.~M. \& {MacAlpine}, G.~M. 1998, \aj, 116, 2624

\bibitem[{{Jiang} {et~al.}(2019){Jiang}, {Stone}, \&
  {Davis}}]{2019ApJ...880...67J}
{Jiang}, Y.-F., {Stone}, J.~M., \& {Davis}, S.~W. 2019, \apj, 880, 67

\bibitem[{{Juod{\v{z}}balis} {et~al.}(2024){Juod{\v{z}}balis}, {Maiolino},
  {Baker}, {Tacchella}, {Scholtz}, {D'Eugenio}, {Witstok}, {Schneider},
  {Trinca}, {Valiante}, {DeCoursey}, {Curti}, {Carniani}, {Chevallard}, {de
  Graaff}, {Arribas}, {Bennett}, {Bourne}, {Bunker}, {Charlot}, {Jiang},
  {Koudmani}, {Perna}, {Robertson}, {Sijacki}, {{\"U}bler}, {Williams}, \&
  {Willott}}]{2024Natur.636..594J}
{Juod{\v{z}}balis}, I., {Maiolino}, R., {Baker}, W.~M., {et~al.} 2024, \nat,
  636, 594

\bibitem[{{Kashino} {et~al.}(2023){Kashino}, {Lilly}, {Matthee}, {Eilers},
  {Mackenzie}, {Bordoloi}, \& {Simcoe}}]{2023ApJ...950...66K}
{Kashino}, D., {Lilly}, S.~J., {Matthee}, J., {et~al.} 2023, \apj, 950, 66

\bibitem[{{Keller} {et~al.}(2023){Keller}, {Munshi}, {Trebitsch}, \&
  {Tremmel}}]{2023ApJ...943L..28K}
{Keller}, B.~W., {Munshi}, F., {Trebitsch}, M., \& {Tremmel}, M. 2023, \apjl,
  943, L28

\bibitem[{{Kocevski} {et~al.}(2024){Kocevski}, {Finkelstein}, {Barro},
  {Taylor}, {Calabr{\`o}}, {Laloux}, {Buchner}, {Trump}, {Leung}, {Yang},
  {Dickinson}, {P{\'e}rez-Gonz{\'a}lez}, {Pacucci}, {Inayoshi}, {Somerville},
  {McGrath}, {Akins}, {Bagley}, {Bisigello}, {Bowler}, {Carnall}, {Casey},
  {Cheng}, {Cleri}, {Costantin}, {Cullen}, {Davis}, {Donnan}, {Dunlop},
  {Ellis}, {Ferguson}, {Fujimoto}, {Fontana}, {Giavalisco}, {Grazian},
  {Grogin}, {Hathi}, {Hirschmann}, {Huertas-Company}, {Holwerda},
  {Illingworth}, {Juneau}, {Kartaltepe}, {Koekemoer}, {Li}, {Lucas}, {Magee},
  {Mason}, {McLeod}, {McLure}, {Napolitano}, {Papovich}, {Pirzkal},
  {Rodighiero}, {Santini}, {Wilkins}, \& {Yung}}]{2024arXiv240403576K}
{Kocevski}, D.~D., {Finkelstein}, S.~L., {Barro}, G., {et~al.} 2024, arXiv
  e-prints, arXiv:2404.03576

\bibitem[{{Kocevski} {et~al.}(2023){Kocevski}, {Onoue}, {Inayoshi}, {Trump},
  {Arrabal Haro}, {Grazian}, {Dickinson}, {Finkelstein}, {Kartaltepe},
  {Hirschmann}, {Aird}, {Holwerda}, {Fujimoto}, {Juneau}, {Amor{\'\i}n},
  {Backhaus}, {Bagley}, {Barro}, {Bell}, {Bisigello}, {Calabr{\`o}}, {Cleri},
  {Cooper}, {Ding}, {Grogin}, {Ho}, {Hutchison}, {Inoue}, {Jiang}, {Jones},
  {Koekemoer}, {Li}, {Li}, {McGrath}, {Molina}, {Papovich},
  {P{\'e}rez-Gonz{\'a}lez}, {Pirzkal}, {Wilkins}, {Yang}, \&
  {Yung}}]{2023ApJ...954L...4K}
{Kocevski}, D.~D., {Onoue}, M., {Inayoshi}, K., {et~al.} 2023, \apjl, 954, L4

\bibitem[{{Kokorev} {et~al.}(2024){Kokorev}, {Caputi}, {Greene}, {Dayal},
  {Trebitsch}, {Cutler}, {Fujimoto}, {Labb{\'e}}, {Miller}, {Iani},
  {Navarro-Carrera}, \& {Rinaldi}}]{2024ApJ...968...38K}
{Kokorev}, V., {Caputi}, K.~I., {Greene}, J.~E., {et~al.} 2024, \apj, 968, 38

\bibitem[{{Komatsu} {et~al.}(2011){Komatsu}, {Smith}, {Dunkley}, {Bennett},
  {Gold}, {Hinshaw}, {Jarosik}, {Larson}, {Nolta}, {Page}, {Spergel},
  {Halpern}, {Hill}, {Kogut}, {Limon}, {Meyer}, {Odegard}, {Tucker}, {Weiland},
  {Wollack}, \& {Wright}}]{Komatsu2011}
{Komatsu}, E., {Smith}, K.~M., {Dunkley}, J., {et~al.} 2011, \apjs, 192, 18

\bibitem[{{Kroupa}(2001)}]{Kroupa2001}
{Kroupa}, P. 2001, \mnras, 322, 231

\bibitem[{{Kubota} \& {Done}(2019)}]{2019MNRAS.489..524K}
{Kubota}, A. \& {Done}, C. 2019, \mnras, 489, 524

\bibitem[{{Labbe} {et~al.}(2025){Labbe}, {Greene}, {Bezanson}, {Fujimoto},
  {Furtak}, {Goulding}, {Matthee}, {Naidu}, {Oesch}, {Atek}, {Brammer},
  {Chemerynska}, {Coe}, {Cutler}, {Dayal}, {Feldmann}, {Franx}, {Glazebrook},
  {Leja}, {Maseda}, {Marchesini}, {Nanayakkara}, {Nelson}, {Pan}, {Papovich},
  {Price}, {Suess}, {Wang}, {Weaver}, {Whitaker}, {Williams}, \&
  {Zitrin}}]{2025ApJ...978...92L}
{Labbe}, I., {Greene}, J.~E., {Bezanson}, R., {et~al.} 2025, \apj, 978, 92

\bibitem[{{Laor} \& {Draine}(1993)}]{1993ApJ...402..441L}
{Laor}, A. \& {Draine}, B.~T. 1993, \apj, 402, 441

\bibitem[{{Laor} \& {Netzer}(1989)}]{Laor1989}
{Laor}, A. \& {Netzer}, H. 1989, \mnras, 238, 897

\bibitem[{{Li} {et~al.}(2024){Li}, {Inayoshi}, {Chen}, {Ichikawa}, \&
  {Ho}}]{2024arXiv240710760L}
{Li}, Z., {Inayoshi}, K., {Chen}, K., {Ichikawa}, K., \& {Ho}, L.~C. 2024,
  arXiv e-prints, arXiv:2407.10760

\bibitem[{{Lin} {et~al.}(2024){Lin}, {Wang}, {Fan}, {Cai}, {Champagne}, {Sun},
  {Volonteri}, {Yang}, {Hennawi}, {Ba{\~n}ados}, {Barth}, {Eilers}, {Farina},
  {Liu}, {Jin}, {Jun}, {Lupi}, {Kakiichi}, {Mazzucchelli}, {Onoue}, {Pan},
  {Pizzati}, {Rojas-Ruiz}, {Schindler}, {Trakhtenbrot}, {Shen}, {Trebitsch},
  {Zhuang}, {Endsley}, {Meyer}, {Li}, {Li}, {Pudoka}, {Tee}, {Wu}, \&
  {Zhang}}]{2024ApJ...974..147L}
{Lin}, X., {Wang}, F., {Fan}, X., {et~al.} 2024, \apj, 974, 147

\bibitem[{{Lupi} {et~al.}(2024){Lupi}, {Trinca}, {Volonteri}, {Dotti}, \&
  {Mazzucchelli}}]{2024A&A...689A.128L}
{Lupi}, A., {Trinca}, A., {Volonteri}, M., {Dotti}, M., \& {Mazzucchelli}, C.
  2024, \aap, 689, A128

\bibitem[{{Lupi} {et~al.}(2022){Lupi}, {Volonteri}, {Decarli}, {Bovino}, \&
  {Silk}}]{2022MNRAS.510.5760L}
{Lupi}, A., {Volonteri}, M., {Decarli}, R., {Bovino}, S., \& {Silk}, J. 2022,
  \mnras, 510, 5760

\bibitem[{{Ma} {et~al.}(2024{\natexlab{a}}){Ma}, {Goulding}, {Greene},
  {Zakamska}, {Wylezalek}, \& {Jiang}}]{Ma2024}
{Ma}, Y., {Goulding}, A., {Greene}, J.~E., {et~al.} 2024{\natexlab{a}}, \apj,
  974, 225

\bibitem[{{Ma} {et~al.}(2024{\natexlab{b}}){Ma}, {Greene}, {Setton},
  {Volonteri}, {Leja}, {Wang}, {Bezanson}, {Brammer}, {Cutler}, {Dayal}, {van
  Dokkum}, {Furtak}, {Glazebrook}, {Goulding}, {de Graaff}, {Kokorev}, {Labbe},
  {Pan}, {Price}, {Weaver}, {Williams}, {Whitaker}, \&
  {Zitrin}}]{2024arXiv241006257M}
{Ma}, Y., {Greene}, J.~E., {Setton}, D.~J., {et~al.} 2024{\natexlab{b}}, arXiv
  e-prints, arXiv:2410.06257

\bibitem[{{Maiolino} {et~al.}(2024{\natexlab{a}}){Maiolino}, {Risaliti},
  {Signorini}, {Trefoloni}, {Juodzbalis}, {Scholtz}, {Uebler}, {D'Eugenio},
  {Carniani}, {Fabian}, {Ji}, {Mazzolari}, {Bertola}, {Brusa}, {Bunker},
  {Charlot}, {Comastri}, {Cresci}, {DeCoursey}, {Egami}, {Fiore}, {Gilli},
  {Perna}, {Tacchella}, \& {Venturi}}]{2024arXiv240500504M}
{Maiolino}, R., {Risaliti}, G., {Signorini}, M., {et~al.} 2024{\natexlab{a}},
  arXiv e-prints, arXiv:2405.00504

\bibitem[{{Maiolino} {et~al.}(2024{\natexlab{b}}){Maiolino}, {Scholtz},
  {Curtis-Lake}, {Carniani}, {Baker}, {de Graaff}, {Tacchella}, {{\"U}bler},
  {D'Eugenio}, {Witstok}, {Curti}, {Arribas}, {Bunker}, {Charlot},
  {Chevallard}, {Eisenstein}, {Egami}, {Ji}, {Jones}, {Lyu}, {Rawle},
  {Robertson}, {Rujopakarn}, {Perna}, {Sun}, {Venturi}, {Williams}, \&
  {Willott}}]{2024A&A...691A.145M}
{Maiolino}, R., {Scholtz}, J., {Curtis-Lake}, E., {et~al.} 2024{\natexlab{b}},
  \aap, 691, A145

\bibitem[{{Maithil} {et~al.}(2024){Maithil}, {Brotherton}, {Shemmer}, {Luo},
  {Du}, {Wang}, {Hu}, {Gallagher}, {Li}, \& {Nemmen}}]{2024MNRAS.528.1542M}
{Maithil}, J., {Brotherton}, M.~S., {Shemmer}, O., {et~al.} 2024, \mnras, 528,
  1542

\bibitem[{{Mathis} {et~al.}(1977){Mathis}, {Rumpl}, \& {Nordsieck}}]{Mathis77}
{Mathis}, J.~S., {Rumpl}, W., \& {Nordsieck}, K.~H. 1977, \apj, 217, 425

\bibitem[{{Matthee} {et~al.}(2024){Matthee}, {Naidu}, {Brammer}, {Chisholm},
  {Eilers}, {Goulding}, {Greene}, {Kashino}, {Labbe}, {Lilly}, {Mackenzie},
  {Oesch}, {Weibel}, {Wuyts}, {Xiao}, {Bordoloi}, {Bouwens}, {van Dokkum},
  {Illingworth}, {Kramarenko}, {Maseda}, {Mason}, {Meyer}, {Nelson}, {Reddy},
  {Shivaei}, {Simcoe}, \& {Yue}}]{2024ApJ...963..129M}
{Matthee}, J., {Naidu}, R.~P., {Brammer}, G., {et~al.} 2024, \apj, 963, 129

\bibitem[{{McKinney} {et~al.}(2012){McKinney}, {Tchekhovskoy}, \&
  {Blandford}}]{McKinney2012}
{McKinney}, J.~C., {Tchekhovskoy}, A., \& {Blandford}, R.~D. 2012, \mnras, 423,
  3083

\bibitem[{{Michel-Dansac} {et~al.}(2020){Michel-Dansac}, {Blaizot}, {Garel},
  {Verhamme}, {Kimm}, \& {Trebitsch}}]{MichelDansac2020}
{Michel-Dansac}, L., {Blaizot}, J., {Garel}, T., {et~al.} 2020, \aap, 635, A154

\bibitem[{{Ni} {et~al.}(2020){Ni}, {Di Matteo}, {Gilli}, {Croft}, {Feng}, \&
  {Norman}}]{2020MNRAS.495.2135N}
{Ni}, Y., {Di Matteo}, T., {Gilli}, R., {et~al.} 2020, \mnras, 495, 2135

\bibitem[{{Oesch} {et~al.}(2023){Oesch}, {Brammer}, {Naidu}, {Bouwens},
  {Chisholm}, {Illingworth}, {Matthee}, {Nelson}, {Qin}, {Reddy}, {Shapley},
  {Shivaei}, {van Dokkum}, {Weibel}, {Whitaker}, {Wuyts}, {Covelo-Paz},
  {Endsley}, {Fudamoto}, {Giovinazzo}, {Herard-Demanche}, {Kerutt},
  {Kramarenko}, {Labbe}, {Leonova}, {Lin}, {Magee}, {Marchesini}, {Maseda},
  {Mason}, {Matharu}, {Meyer}, {Neufeld}, {Prieto Lyon}, {Schaerer}, {Sharma},
  {Shuntov}, {Smit}, {Stefanon}, {Wyithe}, \& {Xiao}}]{2023MNRAS.525.2864O}
{Oesch}, P.~A., {Brammer}, G., {Naidu}, R.~P., {et~al.} 2023, \mnras, 525, 2864

\bibitem[{{Pacucci} \& {Narayan}(2024)}]{2024ApJ...976...96P}
{Pacucci}, F. \& {Narayan}, R. 2024, \apj, 976, 96

\bibitem[{{Pacucci} {et~al.}(2023){Pacucci}, {Nguyen}, {Carniani}, {Maiolino},
  \& {Fan}}]{2023ApJ...957L...3P}
{Pacucci}, F., {Nguyen}, B., {Carniani}, S., {Maiolino}, R., \& {Fan}, X. 2023,
  \apjl, 957, L3

\bibitem[{{P{\'e}rez-Gonz{\'a}lez} {et~al.}(2024){P{\'e}rez-Gonz{\'a}lez},
  {Barro}, {Rieke}, {Lyu}, {Rieke}, {Alberts}, {Williams}, {Hainline}, {Sun},
  {Pusk{\'a}s}, {Annunziatella}, {Baker}, {Bunker}, {Egami}, {Ji}, {Johnson},
  {Robertson}, {Rodr{\'\i}guez Del Pino}, {Rujopakarn}, {Shivaei}, {Tacchella},
  {Willmer}, \& {Willott}}]{2024ApJ...968....4P}
{P{\'e}rez-Gonz{\'a}lez}, P.~G., {Barro}, G., {Rieke}, G.~H., {et~al.} 2024,
  \apj, 968, 4

\bibitem[{{Pfister} {et~al.}(2019){Pfister}, {Volonteri}, {Dubois}, {Dotti}, \&
  {Colpi}}]{2019MNRAS.486..101P}
{Pfister}, H., {Volonteri}, M., {Dubois}, Y., {Dotti}, M., \& {Colpi}, M. 2019,
  \mnras, 486, 101

\bibitem[{{Pouliasis} {et~al.}(2024){Pouliasis}, {Ruiz}, {Georgantopoulos},
  {Vito}, {Gilli}, {Vignali}, {Ueda}, {Koulouridis}, {Akiyama}, {Marchesi},
  {Laloux}, {Nagao}, {Paltani}, {Pierre}, {Toba}, {Habouzit}, {Vijarnwannaluk},
  \& {Garrel}}]{2024A&A...685A..97P}
{Pouliasis}, E., {Ruiz}, A., {Georgantopoulos}, I., {et~al.} 2024, \aap, 685,
  A97

\bibitem[{{Proga}(2005)}]{2005ApJ...630L...9P}
{Proga}, D. 2005, \apjl, 630, L9

\bibitem[{{Rosdahl} {et~al.}(2013){Rosdahl}, {Blaizot}, {Aubert}, {Stranex}, \&
  {Teyssier}}]{Rosdahl2013}
{Rosdahl}, J., {Blaizot}, J., {Aubert}, D., {Stranex}, T., \& {Teyssier}, R.
  2013, \mnras, 436, 2188

\bibitem[{{Rosdahl} \& {Teyssier}(2015)}]{Rosdahl2015}
{Rosdahl}, J. \& {Teyssier}, R. 2015, \mnras, 449, 4380

\bibitem[{{Schneider} {et~al.}(2023){Schneider}, {Valiante}, {Trinca},
  {Graziani}, {Volonteri}, \& {Maiolino}}]{2023MNRAS.526.3250S}
{Schneider}, R., {Valiante}, R., {Trinca}, A., {et~al.} 2023, \mnras, 526, 3250

\bibitem[{{Scholtz} {et~al.}(2023){Scholtz}, {Maiolino}, {D'Eugenio},
  {Curtis-Lake}, {Carniani}, {Charlot}, {Curti}, {Silcock}, {Arribas}, {Baker},
  {Bhatawdekar}, {Boyett}, {Bunker}, {Chevallard}, {Circosta}, {Eisenstein},
  {Hainline}, {Hausen}, {Ji}, {Ji}, {Johnson}, {Kumari}, {Looser}, {Lyu},
  {Maseda}, {Parlanti}, {Perna}, {Rieke}, {Robertson}, {Rodr{\'\i}guez Del
  Pino}, {Sun}, {Tacchella}, {{\"U}bler}, {Venturi}, {Williams}, {Willmer},
  {Willott}, \& {Witstok}}]{2023arXiv231118731S}
{Scholtz}, J., {Maiolino}, R., {D'Eugenio}, F., {et~al.} 2023, arXiv e-prints,
  arXiv:2311.18731

\bibitem[{{Shakura} \& {Sunyaev}(1973)}]{1973A&A....24..337S}
{Shakura}, N.~I. \& {Sunyaev}, R.~A. 1973, \aap, 24, 337

\bibitem[{{Shen} {et~al.}(2020){Shen}, {Hopkins}, {Faucher-Gigu{\`e}re},
  {Alexander}, {Richards}, {Ross}, \& {Hickox}}]{2020MNRAS.495.3252S}
{Shen}, X., {Hopkins}, P.~F., {Faucher-Gigu{\`e}re}, C.-A., {et~al.} 2020,
  \mnras, 495, 3252

\bibitem[{{Silverman} {et~al.}(2023){Silverman}, {Mainieri}, {Ding}, {Liu},
  {Jahnke}, {Hirschmann}, {Kartaltepe}, {Lambrides}, {Onoue}, {Trakhtenbrot},
  {Vardoulaki}, {Bongiorno}, {Casey}, {Civano}, {Faisst}, {Franco}, {Gillman},
  {Gozaliasl}, {Hayward}, {Koekemoer}, {Kokorev}, {Magdis}, {Marchesi}, {Rich},
  {Sparre}, {Suh}, {Tanaka}, \& {Valentino}}]{2023ApJ...951L..41S}
{Silverman}, J.~D., {Mainieri}, V., {Ding}, X., {et~al.} 2023, \apjl, 951, L41

\bibitem[{{Speagle} {et~al.}(2014){Speagle}, {Steinhardt}, {Capak}, \&
  {Silverman}}]{2014ApJS..214...15S}
{Speagle}, J.~S., {Steinhardt}, C.~L., {Capak}, P.~L., \& {Silverman}, J.~D.
  2014, \apjs, 214, 15

\bibitem[{{Stanway} \& {Eldridge}(2018)}]{2018MNRAS.479...75S}
{Stanway}, E.~R. \& {Eldridge}, J.~J. 2018, \mnras, 479, 75

\bibitem[{{Suganuma} {et~al.}(2006){Suganuma}, {Yoshii}, {Kobayashi},
  {Minezaki}, {Enya}, {Tomita}, {Aoki}, {Koshida}, \& {Peterson}}]{Suganuma06}
{Suganuma}, M., {Yoshii}, Y., {Kobayashi}, Y., {et~al.} 2006, \apj, 639, 46

\bibitem[{{Suh} {et~al.}(2024){Suh}, {Scharw{\"a}chter}, {Farina}, {Loiacono},
  {Lanzuisi}, {Hasinger}, {Marchesi}, {Mezcua}, {Decarli}, {Lemaux},
  {Volonteri}, {Civano}, {Yi}, {Han}, {Rawlings}, \&
  {Hung}}]{2024NatAs.tmp..262S}
{Suh}, H., {Scharw{\"a}chter}, J., {Farina}, E.~P., {et~al.} 2024, Nature
  Astronomy [\eprint[arXiv]{2405.05333}]

\bibitem[{{Teyssier}(2002)}]{Teyssier2002}
{Teyssier}, R. 2002, \aap, 385, 337

\bibitem[{{Thomas} {et~al.}(2016){Thomas}, {Groves}, {Sutherland}, {Dopita},
  {Kewley}, \& {Jin}}]{2016ApJ...833..266T}
{Thomas}, A.~D., {Groves}, B.~A., {Sutherland}, R.~S., {et~al.} 2016, \apj,
  833, 266

\bibitem[{{Trebitsch} {et~al.}(2021){Trebitsch}, {Dubois}, {Volonteri},
  {Pfister}, {Cadiou}, {Katz}, {Rosdahl}, {Kimm}, {Pichon}, {Beckmann},
  {Devriendt}, \& {Slyz}}]{Trebitsch2021}
{Trebitsch}, M., {Dubois}, Y., {Volonteri}, M., {et~al.} 2021, \aap, 653, A154

\bibitem[{{Trebitsch} {et~al.}(2019){Trebitsch}, {Volonteri}, \&
  {Dubois}}]{2019MNRAS.487..819T}
{Trebitsch}, M., {Volonteri}, M., \& {Dubois}, Y. 2019, \mnras, 487, 819

\bibitem[{{Trinca} {et~al.}(2022){Trinca}, {Schneider}, {Valiante}, {Graziani},
  {Zappacosta}, \& {Shankar}}]{2022MNRAS.511..616T}
{Trinca}, A., {Schneider}, R., {Valiante}, R., {et~al.} 2022, \mnras, 511, 616

\bibitem[{{Valiante} {et~al.}(2018){Valiante}, {Schneider}, {Zappacosta},
  {Graziani}, {Pezzulli}, \& {Volonteri}}]{2018MNRAS.476..407V}
{Valiante}, R., {Schneider}, R., {Zappacosta}, L., {et~al.} 2018, \mnras, 476,
  407

\bibitem[{{Vito} {et~al.}(2018){Vito}, {Brandt}, {Yang}, {Gilli}, {Luo},
  {Vignali}, {Xue}, {Comastri}, {Koekemoer}, {Lehmer}, {Liu}, {Paolillo},
  {Ranalli}, {Schneider}, {Shemmer}, {Volonteri}, \&
  {Wang}}]{2018MNRAS.473.2378V}
{Vito}, F., {Brandt}, W.~N., {Yang}, G., {et~al.} 2018, \mnras, 473, 2378

\bibitem[{{Volonteri} {et~al.}(2023){Volonteri}, {Habouzit}, \&
  {Colpi}}]{2023MNRAS.521..241V}
{Volonteri}, M., {Habouzit}, M., \& {Colpi}, M. 2023, \mnras, 521, 241

\bibitem[{{Volonteri} {et~al.}(2017){Volonteri}, {Reines}, {Atek}, {Stark}, \&
  {Trebitsch}}]{2017ApJ...849..155V}
{Volonteri}, M., {Reines}, A.~E., {Atek}, H., {Stark}, D.~P., \& {Trebitsch},
  M. 2017, \apj, 849, 155

\bibitem[{{Wasleske} \& {Baldassare}(2024)}]{2024ApJ...971...68W}
{Wasleske}, E.~J. \& {Baldassare}, V.~F. 2024, \apj, 971, 68

\bibitem[{{Weingartner} \& {Draine}(2001)}]{2001ApJ...548..296W}
{Weingartner}, J.~C. \& {Draine}, B.~T. 2001, \apj, 548, 296

\bibitem[{{Yue} {et~al.}(2024){Yue}, {Eilers}, {Ananna}, {Panagiotou}, {Kara},
  \& {Miyaji}}]{2024ApJ...974L..26Y}
{Yue}, M., {Eilers}, A.-C., {Ananna}, T.~T., {et~al.} 2024, \apjl, 974, L26

\end{thebibliography}
%

\appendix

\section{Additional cases}
\begin{figure}[h!]
   \centering
   \includegraphics[width=0.9\columnwidth]{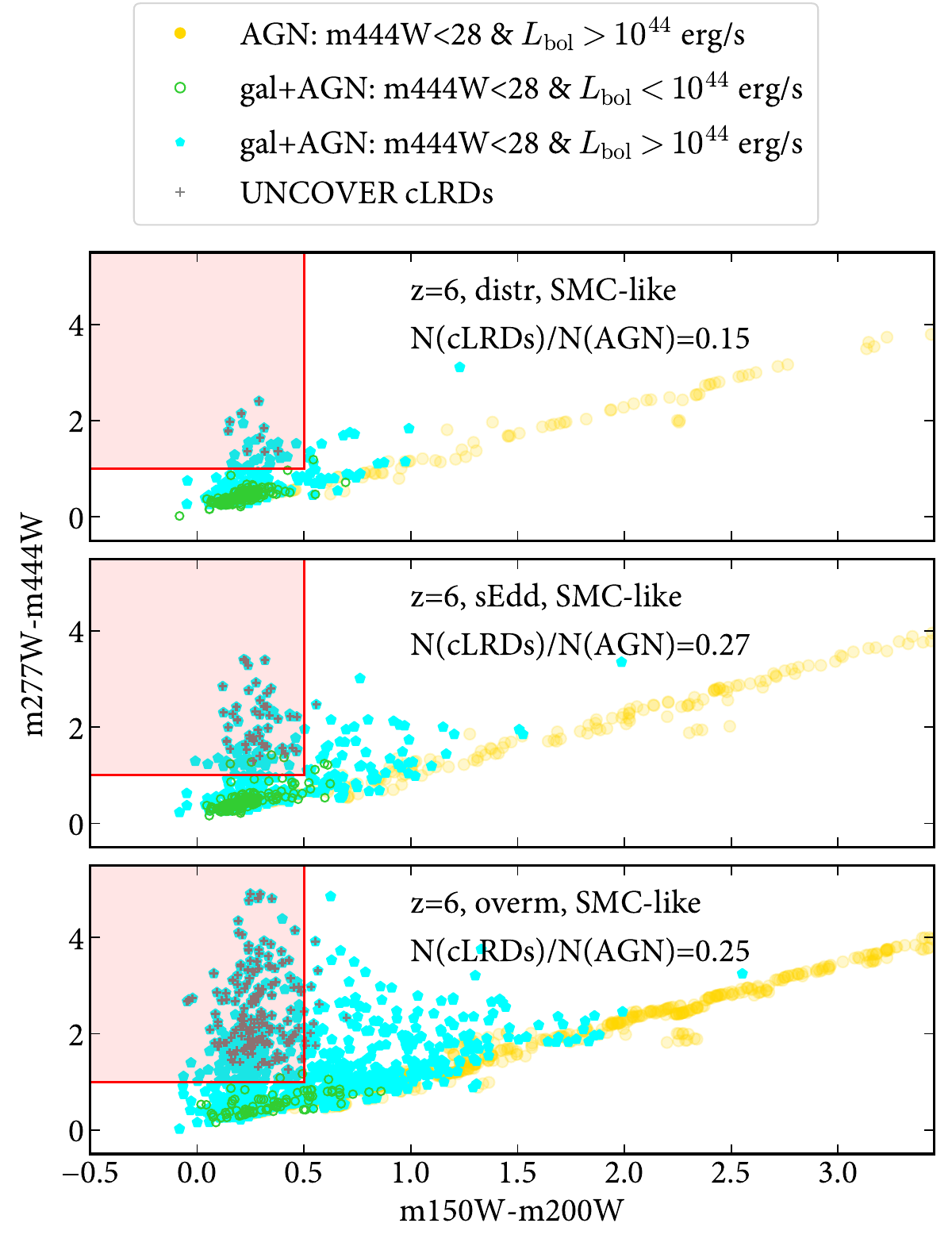}
    \caption{SMILES color selection (shaded red area), with further $L_{\rm bol}>10^{44}\, {\rm erg \, s}^{-1}$ and m444W$<$28 (combined galaxy+AGN). The AGN population is defined as meeting only the $L_{\rm bol}$ and m444W criteria. The yellow points report the colors considering only the AGN emission. The gray crosses highlight the sources that meet the UNCOVER cLRD selection.}
              \label{fig:SMILES}
    \end{figure}

In this Appendix we report how results differ when (i) using different a color selection, or (ii) using an empirical dust law derived for AGN.  We considered how the results differ when considering an alternative color selection of cLRDs following, for instance, \citet{2024ApJ...968....4P},  who adopt m277W$-$m444W$> 1$ and m150W$-$m200W$ < 0.5 $, plus m444W$<28$. We show the results of our models for this selection in Fig~\ref{fig:SMILES}.  This selection is similarly effective, especially in the SMC-like dust case, where more AGN have m150W$-$m200W around zero.

\begin{figure}[h!]
   \centering
   \includegraphics[width=0.9\columnwidth]{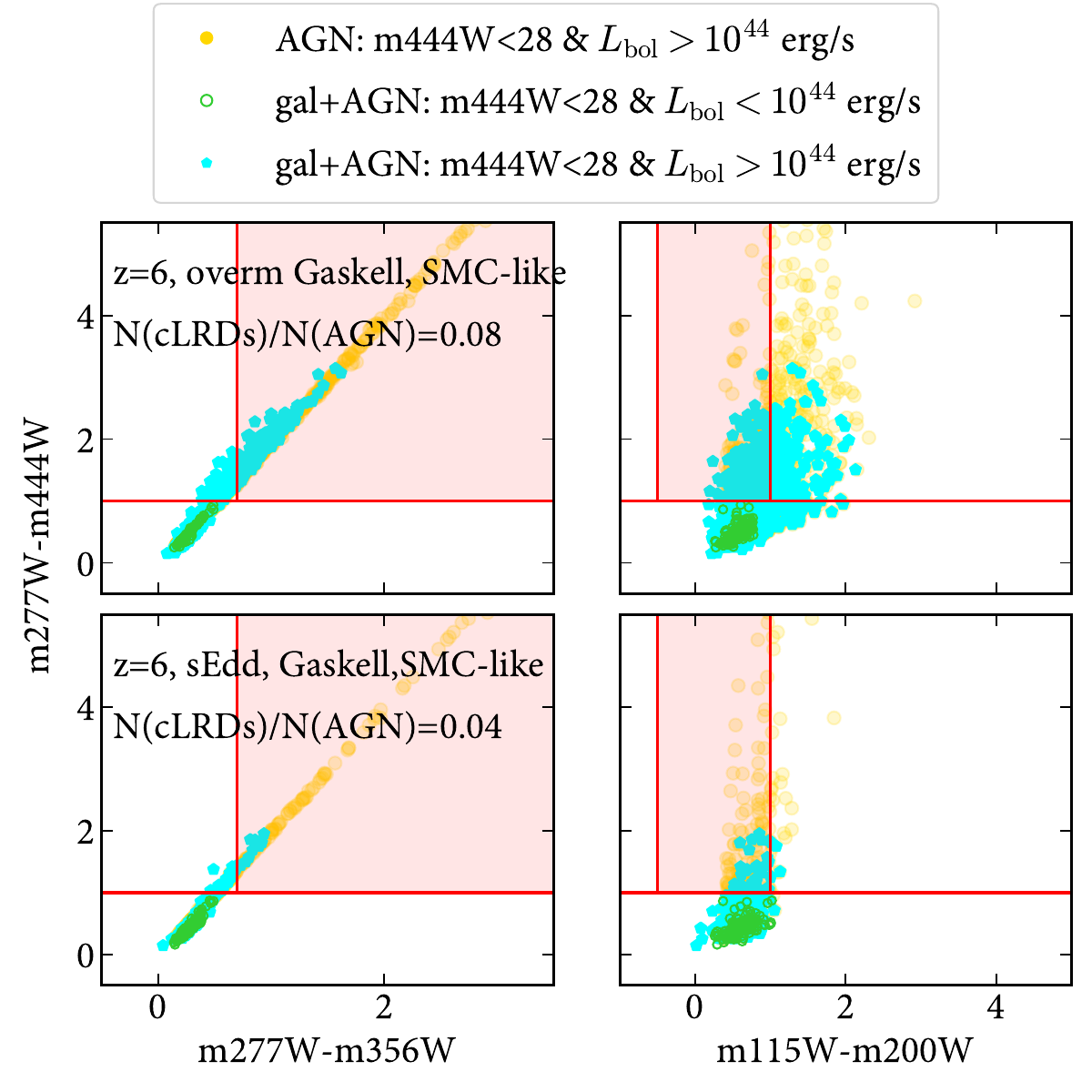}
    \caption{UNCOVER color selection (shaded red area), with further $L_{\rm bol}>10^{44}\, {\rm erg \, s}^{-1}$ and m444W$<$28 (combined galaxy+AGN). The AGN population is defined as meeting only the $L_{\rm bol}$ and m444W criteria. The yellow points report the colors considering only the AGN emission. Here we use the SMC-like extinction curve for the stellar population and the empirical extinction law derived in radio-loud AGN \citep{2004ApJ...616..147G} for the AGN.}
              \label{fig:Gaskell}
    \end{figure}

In Fig.~\ref{fig:Gaskell} we show results using the UNCOVER selection, where we assume that dust for the stellar population is SMC-like and for the AGN is instead described by the empirical extinction curve from \citep{2004ApJ...616..147G}. Whereas \citet{2024arXiv240710760L} find that this extinction curve provides a better fit for little red dots, we find instead that using this curve decreases the fraction of cLRDs among AGN \citep[see also ][for a detailed analysis on the spectrum of an AGN/cLRD]{2024arXiv241006257M}. This is because \citet{2024arXiv240710760L} consider only the AGN contribution, while we include the galaxy contribution as well. In our case it's the stellar population that provides the blue excess, while in the AGN-only case the flatter extinction curve of  \citet{2004ApJ...616..147G} allows one to obtain a blue excess from the AGN only. In fact, if we remove the galaxy contribution to the photometry we find that the fraction of cLRDs among AGN increases to about 10-15\% when using this extinction curve (compared to 4-8\% including the galaxy component, as indicated in the figure).

\end{document}